\documentclass[modern]{aastex62}
\submitjournal{ApJS}
\shorttitle{Globular cluster iron spreads: I. Catalog}
\shortauthors{Bailin}

\newcommand{\feh}{{\ensuremath{\mathrm{[Fe/H]}}}}

\begin{document}
% Title of the paper, and the short title which is used in the headers.
% Keep the title short and informative.
\title{Globular Cluster Intrinsic Iron Abundance Spreads: I. Catalog}

% The list of authors, and the short list which is used in the headers.
% If you need two or more lines of authors, add an extra line using \newauthor
\correspondingauthor{Jeremy Bailin}
\email{jbailin@ua.edu}
\author{Jeremy Bailin}
\affiliation{Department of Physics and Astronomy, University of Alabama Box 870324, Tuscaloosa, AL, 35487-0324, USA}

% These dates will be filled out by the publisher

% Don't change these lines

\begin{abstract}
We present an up-to-date catalog of intrinsic iron abundance spreads in the 55 Milky Way globular clusters for which sufficiently precise spectroscopic measurements are available. Our method combines multiple datasets when possible to improve the statistics, taking into account the fact that different methods and instruments can lead to systematically offset metallicities. Only high spectral resolution ($R>14,000$) studies that measure the equivalent widths of individual iron lines are found to have uncertainties on the metallicities of the individual stars that can be calibrated sufficiently well for the intrinsic dispersion to be separated cleanly from random measurement error. The median intrinsic iron spread is found to be 0.045~dex, which is small but unambiguously measured to be non-zero in most cases. There is large variation between clusters, but more luminous globular clusters, above $10^5~L_{\odot}$, have increasingly large iron spreads on average; no trend between iron spread and metallicity is found.
\end{abstract}



%%%%%%%%%%%%%%%%%%%%%%%%%%%%%%%%%%%%%%%%%%%%%%%%%%

%%%%%%%%%%%%%%%%% BODY OF PAPER %%%%%%%%%%%%%%%%%%

\section{Introduction}

Globular clusters (GCs) were once thought of as single stellar populations, whose stars all share the same age and global composition. Better observations have radically reshaped that picture -- it is now clear that most, if not all, GCs contain multiple populations with different abundances of many intermediate-mass elements (e.g. O, Na, Mg, Al; see \citealp{BastianLardo18} for a recent review) and in some cases somewhat different ages. However, the products of supernovae, as traced by global metallicity and in particular the abundance of iron, is still considered to be uniform amongst the stars in a GC in most cases. A few GCs (e.g. M~54, $\omega$ Cen, and a handful of others) are generally accepted to have iron abundance spreads, but these clusters are considered ``anomalous'' and unrepresentative of typical GCs \citep[e.g.][]{Marino15-N5286}.

However, more careful examination of the data suggests that while iron abundance dispersions are small in most GCs, they are measurable \citep{WillmanStrader12,Leaman12,Bailin18}.
A spread in iron abundances could be an important clue to the origin of GCs. Self-enrichment in GCs is required to explain the large variations and correlations in the abundances of intermediate-mass elements  and, in some models of self-enrichment, can also lead to iron abundance spreads \citep{Leaman12,Renzini13,Bailin18}. Alternatively, iron abundance spreads have often been taken as a sign that the object is actually the stripped remnant of a dwarf galaxy, all of which have significant metallicity spreads, rather than a true GC \citep{Bekki03,Carretta10-M54,Massari14-T5,Marino15-N5286}. Understanding how iron abundances fit into the origins of GCs requires a good homogeneously-produced catalog of abundance spreads in a large sample of GCs.

However, such homogeneity has not existed until now. There are three major problems that have made this difficult:
\begin{enumerate}
  \item Sample sizes have often been small because it was difficult to obtain sufficiently high signal-to-noise data with high spectral resolution on a large number of stars before the existence of multi-object spectrographs on 8m class telescopes like VLT FLAMES and Magellan M2FS.
 \item The intrinsic dispersions are often comparable to or smaller than the uncertainty on each individual measurement.
 \item Relatedly, the results are quite sensitive to the assumed uncertainty on each individual measurement, which is difficult to estimate correctly.
 \item There are a number of significant sources of systematic uncertainty, including model atmospheres and non-LTE and 3D effects, that prevent combining multiple studies to increase sample sizes.
\end{enumerate}
There has been no particular pattern to how authors quote their abundance spreads; some quote the observed rms spread while others remove a typical measurement uncertainty in quadrature, and rarely are the spreads themselves given error estimates. When it comes to the uncertainties on abundance measurements of individual stars, some authors combine random and systematic uncertainties (the former inflate abundance spreads while the latter do not), some quote the line-to-line measurement rms of the stellar absorption lines, some quote the line-to-line rms divided by the square root of the number of lines, and some quote the sensitivity of the metallicity measurements to model atmosphere parameters, or various combinations of these possibly added in quadrature. Although there are reasonable justifications for many of these choices, the diversity severely complicates a direct comparison between studies.

As a first step towards using metallicity spreads within GCs to understand their origins, this work collects all of the available data and reanalyzes it homogeneously, presenting an up-to-date homogeneous catalog of iron abundance spread measurements in GCs.

In Section~\ref{sec:literature}, we overview the literature that has been used to generate this catalog. Section~\ref{sec:analysis} contains a description of the method used to determine the intrinsic metallicity dispersions from the available data, including a method to combine multiple datasets. The results, including cross-validation tests for different datasets, are presented in Section~\ref{sec:results}, and conclusions are presented in Section~\ref{sec:conclusions}.

\section{Literature}\label{sec:literature}
The last catalog of GC intrinsic dispersions was performed by \citet{WillmanStrader12}; see also \citet{Leaman12}. The criteria we adopt for which studies to include, which is similar to those used by \citet{WillmanStrader12}, are:
\begin{enumerate}
 \item Metallicities must be based on spectroscopic measurements of iron lines. Although there is a wealth of information from using other metallicity indicators, such as  photometry or calcium triplet spectroscopy, a homogeneous catalog requires that all studies are measuring the same aspect of the cluster stars. 
 \item The iron abundance must be based on equivalent width measurements of individual iron lines, not fitting based on synthetic spectra, whose uncertainty is much harder to quantify (see Section~\ref{sec:xvalid}). In practice, this corresponds to a spectral resolution lower limit of $R>14,000$.
 \item There must be at least 5 stars per dataset (although not all such datasets are usable if the total amount of data from all datasets is not sufficient; see Section~\ref{sec:combimethod} and Figure~\ref{fig:Ntest}).
 \item All observations should be of red giant branch (RGB) stars -- no asymptotic giant branch (AGB) stars, horizontal branch/red clump stars, subgiants, or main sequence stars are used. RGB stars and AGB stars are the brightest stars in an old stellar population, so the most likely to be observed spectroscopically, but have different atmospheres and so their temperature and surface gravity scales may differ; moreover, the current generation of AGB model atmospheres appear to have fundamental issues that prevent consistent abundance measurements \citep{Mucciarelli19}.
 The relatively small range in effective temperature and surface gravity on the upper RGB dramatically reduces the likelihood of artificial inflation of abundance spreads by any residual systematic trend between them and abundance. Observations that span a range of evolutionary stages find that there can be systematic differences in derived metallicities, for example between main sequence turnoff stars, subgiants, and red giants \citep{Nordlander12,Husser16}, which might be due to systematic discrepancies in stellar atmosphere parameters or models \citep{Spite16} or physical settling of elements in the atmospheres at different evolutionary stages \citep{Nordlander12}. Regardless of the cause, these cannot be true differences between the global composition of the stars, which means that including multiple evolutionary stages will artificially inflate the dispersion.
 We have used the evolutionary stage assigned to each star by the study authors unless there was specific evidence, such as variability, that a giant was an AGB star. It is possible that the tip of the RGB could be contaminated by a few AGB stars, but in the examples we tested we saw no difference in our results from excluding the brightest stars.
\end{enumerate}

We began with the \citet{WillmanStrader12} catalog as our base.
We then went methodically through every cluster listed in the GC catalog of \citet{BalbinotGieles18}, which is an augmented version of the 2010 version of the Harris
catalogue \citep{Harris96,Harris10}, and performed a literature search on NASA's Astrophysics Data System\footnote{\url{http://adsabs.harvard.edu}} for any papers from 2012 or later with the word ``abundance'' or ``metallicity'' in the abstract and a match for the object in question. Earlier data sources not used by \citet{WillmanStrader12} were also included when they came to our attention, and we went back to the original data that \citet{WillmanStrader12} used when it was possible to combine it with other data.

\section{Analysis}\label{sec:analysis}

\subsection{Abundance Measurements}
In the literature, there are many variations on the methodology of turning absorption line equivalent width measurements into metallicities. Some studies calculate \feh\ independently from the \ion{Fe}{1} and \ion{Fe}{2} lines, while others either average them or force them to be equal to each other to enforce ionization equilibrium and only report a single \feh\ value for each star. Although \ion{Fe}{1} can be subject to non-LTE systematic shifts due to the small fraction of iron that is neutral at stellar atmospheric temperatures, the significantly larger number of \ion{Fe}{1} lines than \ion{Fe}{2} lines and better-determined oscillator strengths mean that the \ion{Fe}{1} abundances are measured more precisely. Therefore, for the goal of determining the spread in metallicity, which requires precision but is insensitive to systematic shifts, we adopt the \ion{Fe}{1} abundances when available, otherwise we adopt the reported \feh.

\subsection{Abundance Uncertainties}
The dispersion in \feh\  within a cluster is small and of similar scale to the uncertainties in the abundance measurements of the individual stars. Therefore, having a proper understanding of those uncertainties is critical for determining how much of the observed star-to-star scatter is due to random measurement error vs. intrinsic spread in abundances. Unfortunately, accurately estimating the uncertainty is difficult, and different authors have often chosen to quote different quantities when estimating their magnitude.

It is critical to separate random uncertainties from systematic uncertainties. Random uncertainties that operate independently on each star increase the measured star-to-star variation in metallicity; major sources of random uncertainty include photon statistics, continuum placement, and the statistics from the number of absorption lines that are used. On the other hand, uncertainties that systematically shift measured metallicities higher or lower change global measured metallicities but do not increase star-to-star dispersion; major sources of systematic uncertainty include instrumental effects, stellar atmosphere models, line lists, oscillator strengths, non-LTE and 3D effects, and details of how the analysis is performed.

Atmospheric parameters (effective temperature $T_{\mathrm{eff}}$, surface gravity $\log g$, and microturbulent velocity) can have both random and systematic effects: derived metallicities are sensitive to the parameters, and different methods often derive systematically higher or lower values of these parameters relative to each other, and consequently derive systematically shifted metallicities; on the other hand, if there are random errors in these parameters (e.g. due to photometric uncertainty for photometrically-derived $T_{\mathrm{eff}}$ and $\log g$), then those random errors propagate into random errors in the derived metallicity. Most authors calculate the sensitivity of abundances to the atmospheric parameters and use that to translate an estimated error on atmospheric parameters into an estimated error on metallicity. However, the estimation of atmospheric parameter uncertainties is itself heterogenous: some studies measure a dispersion and therefore accurately derive the random uncertainty, some look at total offsets that include both random and systematic components, and some use typical values from the literature that might or might not be relevant for the study in question.
When it comes to translating the atmospheric parameter uncertainty into abundance uncertainty,
many studies assume that the sensitivity to each parameter acts independently as a random component, when in fact this is a major overestimate both because some of this variation is systematic, but more importantly because there are large covariances between the effects of different atmospheric parameters (especially between $T_{\mathrm{eff}}$ and $\log g$; \citealp{McWilliam95}).

Whenever possible, we have adopted empirical measurements of the random component of the uncertainty. Most often, this corresponds to $\sigma_{\mathrm{lines}}/\sqrt{N_{\mathrm{lines}}}$, where $\sigma_{\mathrm{lines}}$ is the rms dispersion in the measurements of \feh\ from the different \ion{Fe}{1} lines and $N_{\mathrm{lines}}$ is the number of \ion{Fe}{1} lines used (as in \citealp{WillmanStrader12}; J. Strader, priv. comm.). Uncertainties based on sensitivity to atmospheric parameters are included when the studies properly take covariances into account and take care about how the atmospheric parameter uncertainties are estimated, otherwise the line-to-line component is a better estimate due to the issues discussed above.
Studies whose adopted uncertainties deserve special
note are listed in Section~\ref{sec:individual-notes}.

In order to identify cases where the abundance errors have been overestimated, we have looked at the rms $z$-scores, $\sigma_z$. Assuming that abundance errors have a normal distribution, then if we define $z_i$ to be the difference between a datapoint $x_i$ and the mean of the dataset $\bar{x}$ in units of the measurement uncertainty for that datapoint $e_i$,
\begin{equation}
z_i = \frac{x_i - \bar{x}}{e_i},
\end{equation}
then for a cluster with no intrinsic dispersion, the values $z_i$ should follow a normal distribution of width $\sigma_z=1$. Intrinsic dispersion can only increase $\sigma_z$. Therefore, \textbf{if the nominal errors $e_i$ are accurate estimates of the random uncertainty in the data}, $\sigma_z \ge 1$. The only way to have $\sigma_z \ll 1$ is if the errors have been overestimated.

\subsection{Measuring Intrinsic Dispersion with Combined Datasets}\label{sec:combimethod}

\begin{figure}
\plotone{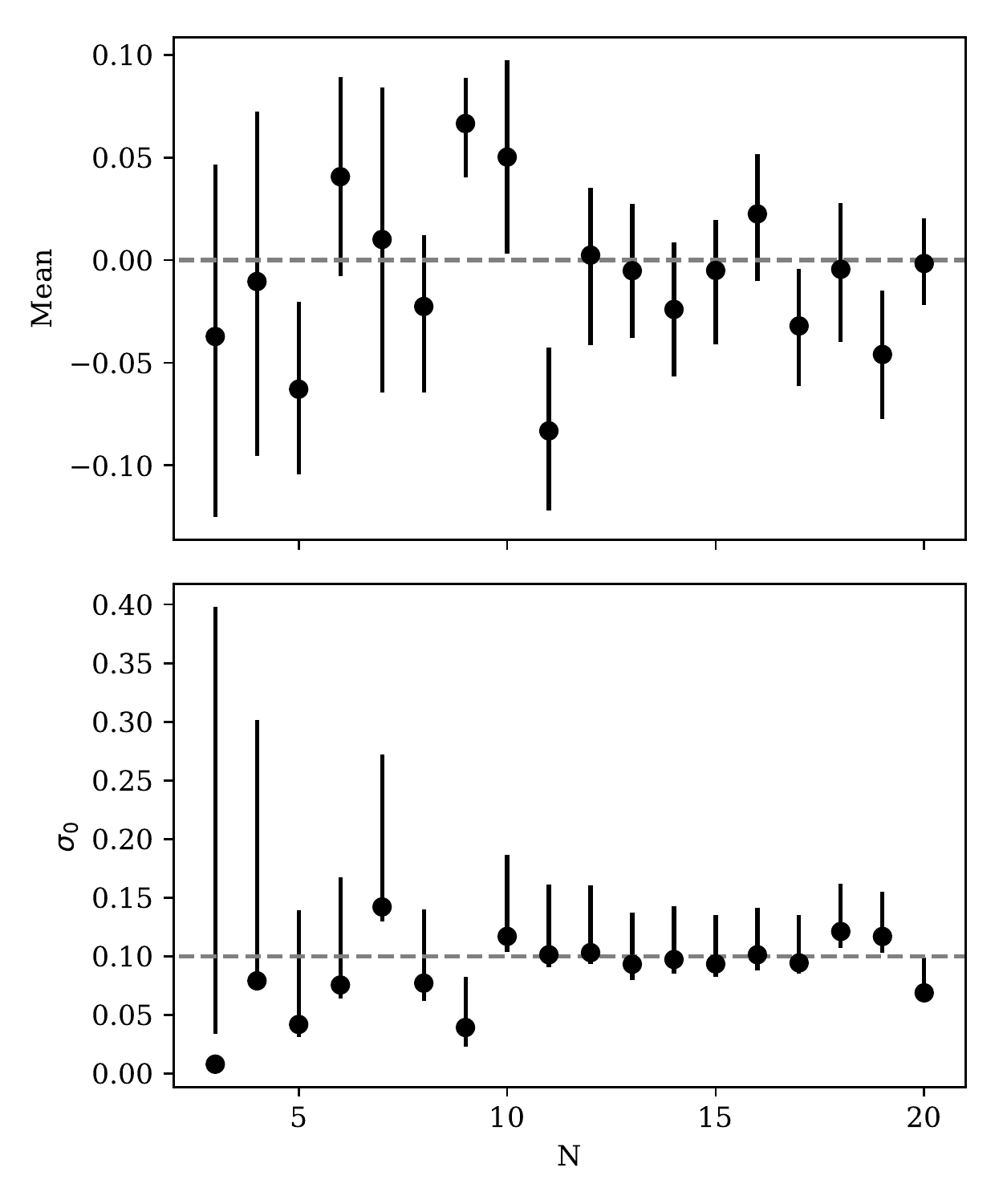}
\caption{Tests of how well the mean (top) and width $\sigma_0$ (bottom) are recovered by the algorithm for random Gaussian datasets with different values of $N$. In all cases, the mean was 0 and $\sigma_0=0.1$ with a random error per measurement of $e=0.05$, typical values for the datasets we use. $\sigma_0$ is recovered to within the derived error bars for $N \ge 10$.\label{fig:Ntest}}
\end{figure}

For many clusters, the number of stars in any individual study is small, often a dozen or fewer. To determine the minimum amount of data required, we have generated random Gaussian datasets of various size $N$ with an intrinsic $\sigma_0=0.1$, a mean of 0, and random errors per measurement of $e=0.05$ for all measurements; this ratio of $\sigma_0/e_i$ is typical for the actual datasets analyzed. Figure~\ref{fig:Ntest} shows the recovered mean and values of $\sigma_0$ as a function of $N$; $\sigma_0$ is recovered to within the derived error bars for $N \ge 10$, although the error bars remain large and more data is much better. We therefore require that the total number of datapoints be $N \ge 10$.

\begin{figure}
\plottwo{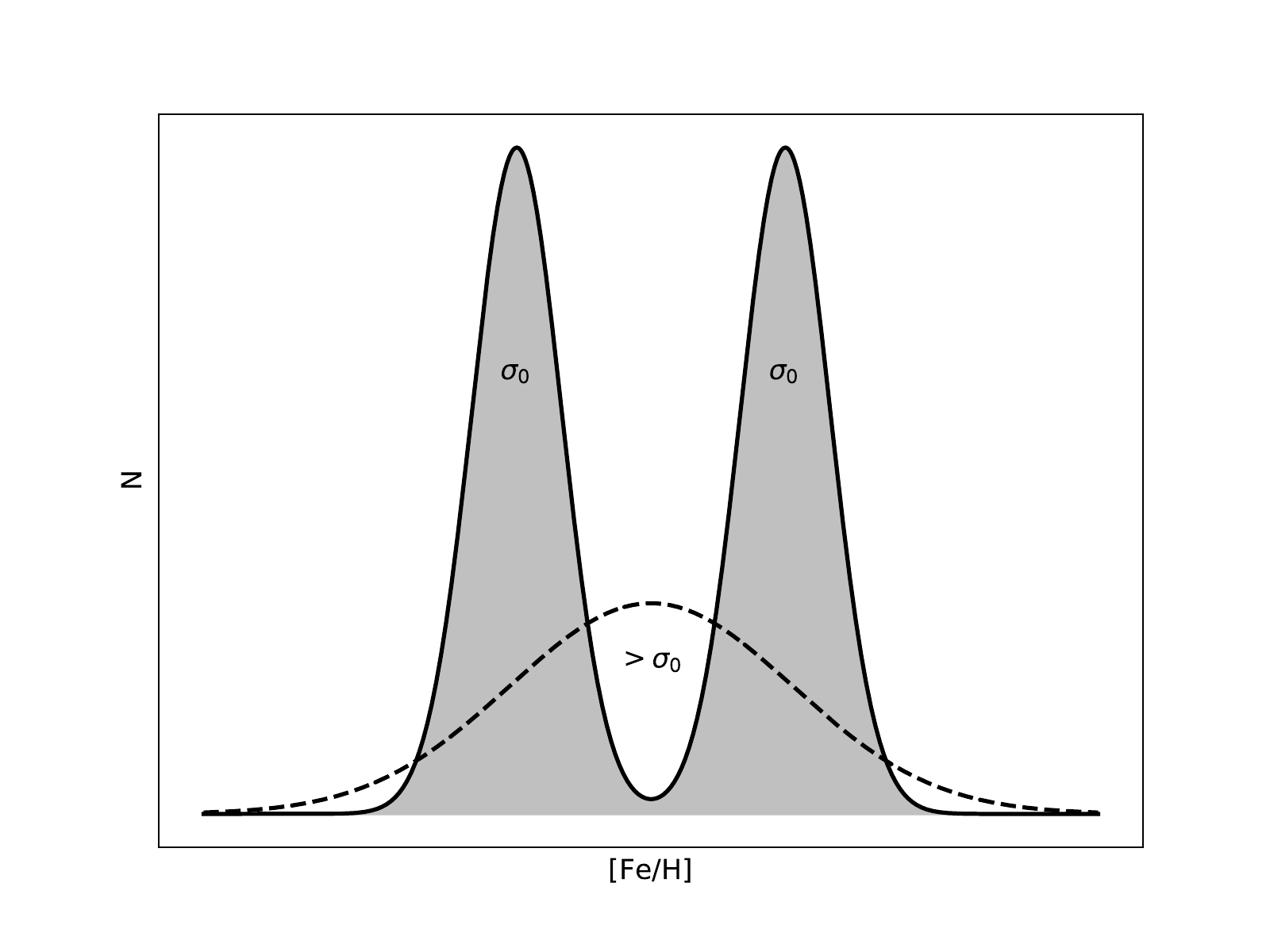}{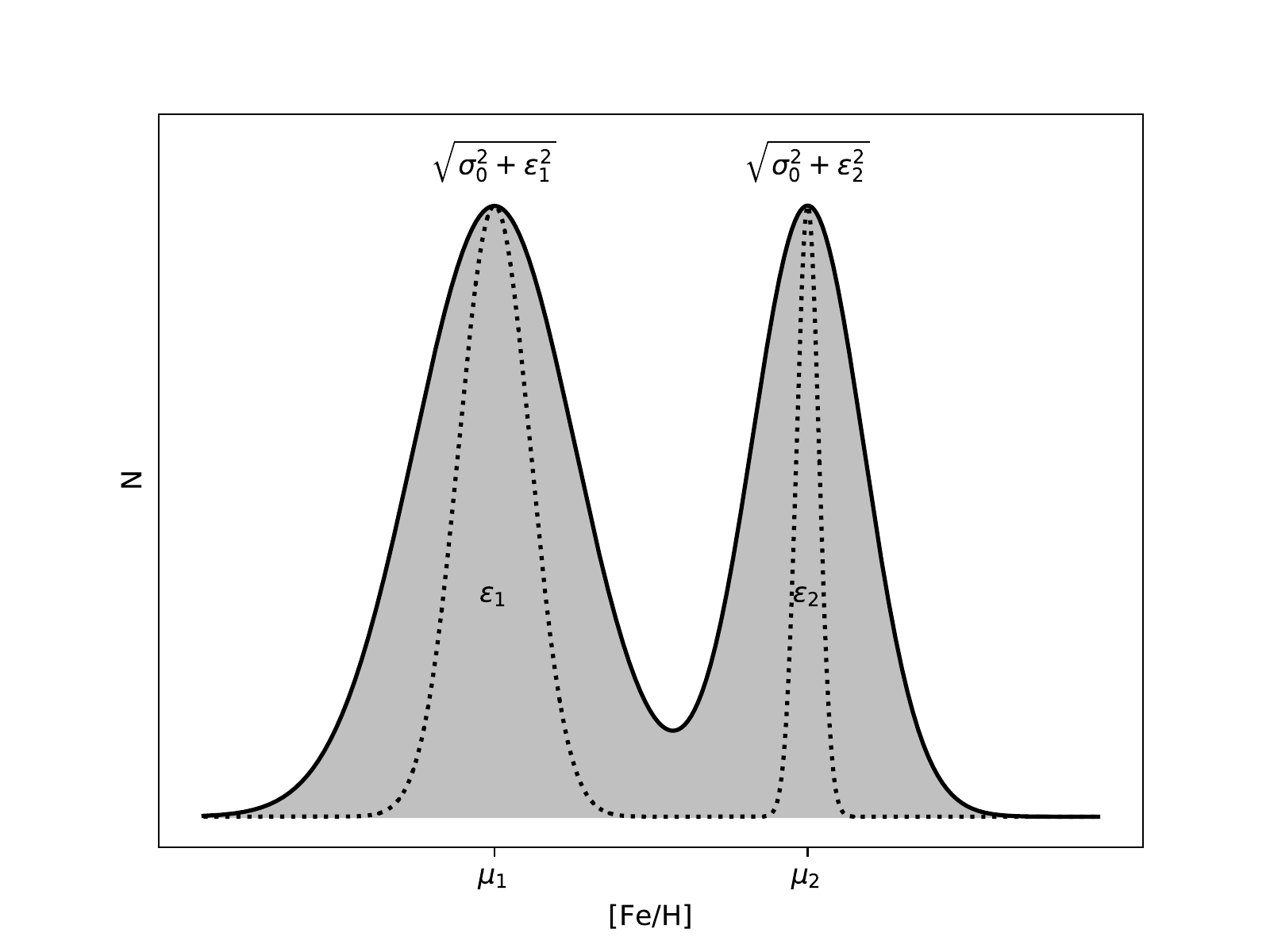}
 \caption{\label{fig:sigma-inflation}\label{fig:simultaneous-sigma}%
 	Left: Schematic view of how combining two different data sets with the same intrinsic width $\sigma_0$ but a systematic offset in the mean value (solid curves) can lead to an artificially inflated measurement of the width (dashed curve).
 	Right: Schematic view of how we simultaneously fit multiple data sets at the same time. For example, if all of the stars in dataset 1 have a random uncertainty $\epsilon_1$ and a mean $\mu_1$, and all of the stars in dataset 2 have a random uncertainty $\epsilon_2$ and a mean $\mu_2$, then we can simultaneously fit for the three parameters $\sigma_0$, $\mu_1$, and $\mu_2$ without biasing the measurement of $\sigma_0$.%
 	}
\end{figure}

In order to increase the amount of data, and therefore obtain small error bars, we have performed a combined analysis of multiple datasets when they are available to increase the sample size. However, when doing so we must account for systematic effects.
Different studies can find systematically offset values of \feh\ for many reasons, including choices of model atmospheres, spectral region, instrumental effects, spectral resolution, treatment of non-LTE effects, and use of only \ion{Fe}{1} lines versus both \ion{Fe}{1} and \ion{Fe}{2} lines. Combining multiple datasets with systematic offsets can lead to inflated values of $\sigma_0$ (Figure~\ref{fig:sigma-inflation}).

In order to prevent such artificially inflated measurements,
we have adopted a Bayesian Markov Chain Monte Carlo (MCMC) approach to fitting multiple datasets simultaneously to a normal distribution (Figure~\ref{fig:simultaneous-sigma}). For a single dataset $i$ with $1 \le j \le N_i$ data points  $\{x_{i,j}\}$ and uncertainties $\{e_{i,j}\}$, the likelihood 
of a normal distribution with central value $\mu_i$ and width $\sigma_0$ is
\begin{equation}
p_i(\{x_{i,j}\} | \mu_i, \sigma_0) = \prod_{j=1}^{N_i}
\frac{1}{\sqrt{2\pi} (\sigma_0^2 + e_{i,j}^2)^{1/2}} \exp\left(-\frac{1}{2} \frac{(x_{i,j}-\mu_i)^2}{ (\sigma_0^2 + e_{i,j}^2)}\right).
\end{equation}
If there are $M$ data sets with identical values of $\sigma_0$ but possibly offset central values $\mu_i$, the likelihood of the full data is
\begin{equation}
p(\{x_{i,j}\} | \mu_{1\ldots M}, \sigma_0) = \prod_{i=1}^{M} p_i(\{x_{i,j}\} | \mu_i, \sigma_0),
\end{equation}
and for flat priors on the $M+1$ parameters $\left\{\sigma_0, \mu_{1\ldots M}\right\}$, the logarithm of the posterior pdf is
\begin{equation}
L_p = \textrm{const} - \frac{1}{2} \sum_{i=1}^{M} \sum_{j=1}^{N_i} \left( \ln(\sigma_0^2 + e_{i,j}^2) + \frac{(x_{i,j} - \mu_i)^2}{\sigma_0^2 + e_{i,j}^2} \right).
\end{equation}
We have used the \texttt{emcee} package \citep{emcee} to find the best fit values and confidence intervals for the data, with uniform priors on $\mu_{1\ldots M}$ and a uniform prior on $\sigma_0 \ge 0$. With $M=1$ dataset, this is the same as the method used by \citet{WillmanStrader12}, and so in cases where there are no new additional datasets, we adopt their values.

This approach to fitting multiple datasets has a number of significant advantages. First, it increases the sample size of stars allowing for much more precise measurements of $\sigma_0$, and for measurements in cases where no individual study contains enough stars. Second, it makes the results more robust to idiosyncrasies in an individual dataset. Third, even when different datasets contain observations of the same stars, which would artificially reduce the measured $\sigma_0$ if all data were simply analyzed together due to the presence of non-independent data, our method compares each datapoint only to other datapoints in the same dataset, not to the data in different datasets, and so does not suffer from this effect.

Although this technique allows multiple studies to be analyzed simultaneously, each dataset must have enough information for an independent determination of the central value $\mu_i$ and constraints on $\sigma_0$; therefore, individual datasets with $N \le 4$ stars have not been considered. As discussed earlier, a single dataset with at least 10 data points is required; similarly, for multiple datasets the total number of datapoints minus the number of additional studies (to account for the additional degree of freedom $\mu_i$ contributed by each additional dataset) must be greater than 10. For single studies that combine data taken with multiple instruments, the data from each instrument is treated as a separate dataset due to the possible instrumental systematic effects.

The MCMC algorithm has been run on each dataset individually as well as on the combined dataset, which is useful for cross-validation and identifying cases where studies disagree and further investigation is warranted; these are noted in Section~\ref{sec:individual-notes}.

We have used the MCMC chains to calculate the overall mean value of \feh\ for each cluster. For each sample in the chain, we calculate an average of the central values $\mu_i$ of the datasets each weighted by the number of stars $N_i$, and use the median and 68\%\ confidence interval of these samples as the estimate of the mean \feh\ of the cluster. This estimate is not as homogeneous as the estimates of $\sigma_0$ because it is affected by systematic offsets between studies and non-LTE effects in the \ion{Fe}{1} lines.

\begin{figure}
\plotone{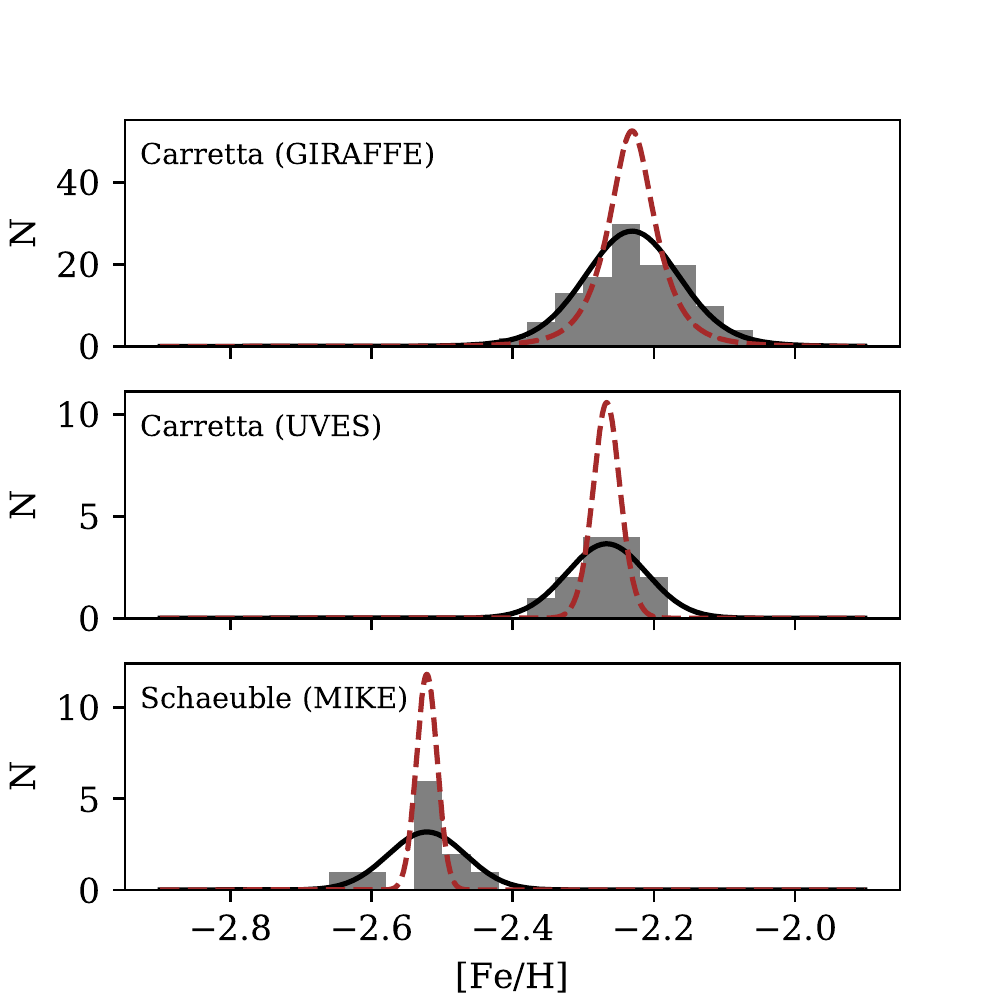}
\caption{Metallicity distribution function for NGC~4590, based on \citet{Carretta09-GIRAFFE} GIRAFFE data (top), \citet{Carretta09-UVES} UVES data (middle), and \citet{Schaeuble15} Magellan MIKE data (bottom). Brown dashed curves denote the expected distribution if there were no intrinsic dispersion, based on the magnitude of the random errors for each star in each dataset, while the black solid curve is the combined effect of the random errors and an intrinsic Gaussian of width $\sigma_0=0.053$~dex, as derived by the MCMC algorithm. The algorithm simultaneously fits the larger-$N$ but lower-resolution GIRAFFE data and the higher-resolution UVES and MIKE data with a consistent value of $\sigma_0$, while not artificially inflating it because of the systematic offset in \feh\ between studies.%
\label{fig:samplehist}}
\end{figure}

An example is shown in Figure~\ref{fig:samplehist} for NGC~4590, which uses three datasets. The higher resolution UVES and MIKE data have smaller random errors and therefore a slightly narrower spread in measured \feh, while the somewhat lower resolution GIRAFFE data points each have larger random errors, but the larger number of stars provides statistical power. Moreover, \citet{Schaeuble15} find systematically lower values of \feh\ across the board. The MCMC algorithm is able to come up with a value of $\sigma_0$ that is simultaneously consistent with all three datasets, and is unaffected by the systematic shift between them.

The quoted error bars on $\sigma_0$ are \textbf{only} random errors derived from the MCMC fitting procedure. The lower error bars tend to be much smaller than the upper error bars (e.g. Figure~\ref{fig:Ntest}); in other words, the preferred value is essentially the smallest $\sigma_0$ that is consistent with the data.
The quoted error bars  are likely smaller than systematic errors due to the choice of studies to include and the assumption of a Gaussian distribution. However, relative comparison of different clusters should be valid, and the robustness of the method is demonstrated in Section~\ref{sec:xvalid}.

\section{Results}\label{sec:results}

\subsection{Catalog}

\startlongtable
\begin{deluxetable}{llccc}
\tablecaption{Derived dispersions $\sigma_0$ and average metallicity \feh\ for each cluster.\label{table:sigma0}}

\tablehead{ \colhead{Cluster Name} & \colhead{Alt. Name} & \colhead{[Fe/H]} & \colhead{$\sigma_0$} & \colhead{References} }
\startdata
NGC 104 & 47 Tuc & $-0.747\pm 0.003$ & $0.033^{+0.003}_{-0.002}$ & 1,2,3 \\
NGC 288 &  & $-1.226\pm 0.004$ & $0.037^{+0.004}_{-0.003}$ & 1,2 \\
NGC 362 &  & $-1.213\pm 0.024$ & $0.074^{+0.030}_{-0.008}$ & 4 \\
NGC 1851 &  & $-1.157\pm 0.005$ & $0.046^{+0.004}_{-0.003}$ & 5 \\
NGC 1904 & M 79 & $-1.550\pm 0.005$ & $0.027^{+0.006}_{-0.002}$ & 1,2 \\
NGC 2419 &  & $-2.095\pm 0.019$ & $0.032^{+0.013}_{-0.009}$ & 5 \\
NGC 2808 &  & $-1.120\pm 0.003$ & $0.035^{+0.003}_{-0.002}$ & 6,7 \\
NGC 3201 &  & $-1.496\pm 0.004$ & $0.044^{+0.004}_{-0.002}$ & 1,2 \\
NGC 4590 & M 68 & $-2.255\pm 0.006$ & $0.053^{+0.008}_{-0.002}$ & 1,2,8 \\
NGC 4833 &  & $-2.070\pm 0.003$ & $0.013^{+0.004}_{-0.002}$ & 9,10 \\
NGC 5024 & M 53 & $-1.995\pm 0.011$ & $0.071^{+0.008}_{-0.007}$ & 11,12 \\
NGC 5053 &  & $-2.450\pm 0.014$ & $0.041^{+0.018}_{-0.004}$ & 13 \\
NGC 5139 & $\omega$ Cen & $-1.647\pm 0.009$ & $0.271^{+0.007}_{-0.007}$ & 5 \\
NGC 5272 & M 3 & $-1.391\pm 0.012$ & $0.097^{+0.012}_{-0.007}$ & 12 \\
NGC 5286 &  & $-1.727\pm 0.014$ & $0.103^{+0.015}_{-0.006}$ & 14 \\
NGC 5466 &  & $-1.865\pm 0.027$ & $<0.075$ & 12 \\
NGC 5634 &  & $-1.869\pm 0.015$ & $0.081^{+0.016}_{-0.008}$ & 15 \\
NGC 5694 &  & $-2.017\pm 0.012$ & $0.046^{+0.015}_{-0.006}$ & 16,17 \\
NGC 5824 &  & $-2.174\pm 0.006$ & $0.058^{+0.006}_{-0.003}$ & 18,19 \\
NGC 5904 & M 5 & $-1.259\pm 0.003$ & $0.041^{+0.005}_{-0.001}$ & 1,2,12 \\
NGC 5986 &  & $-1.527\pm 0.018$ & $0.061^{+0.023}_{-0.005}$ & 20 \\
NGC 6093 & M 80 & $-1.789\pm 0.003$ & $0.014^{+0.003}_{-0.002}$ & 21 \\
NGC 6121 & M 4 & $-1.166\pm 0.004$ & $0.050^{+0.005}_{-0.001}$ & 1,2,3 \\
NGC 6139 &  & $-1.593\pm 0.006$ & $0.033^{+0.006}_{-0.004}$ & 22 \\
NGC 6171 & M 107 & $-0.949\pm 0.008$ & $0.047^{+0.010}_{-0.003}$ & 2,12 \\
NGC 6205 & M 13 & $-1.443\pm 0.014$ & $0.101^{+0.014}_{-0.009}$ & 12 \\
NGC 6218 & M 12 & $-1.315\pm 0.004$ & $0.029^{+0.004}_{-0.002}$ & 1,23 \\
NGC 6229 &  & $-1.129\pm 0.020$ & $0.044^{+0.032}_{-0.012}$ & 24 \\
NGC 6254 & M 10 & $-1.559\pm 0.005$ & $0.049^{+0.004}_{-0.003}$ & 1,2 \\
NGC 6266 & M 62 & $-1.075\pm 0.013$ & $0.041^{+0.015}_{-0.005}$ & 25 \\
NGC 6273 & M 19 & $-1.612\pm 0.022$ & $0.161^{+0.020}_{-0.015}$ & 26,27 \\
NGC 6341 & M 92 & $-2.239\pm 0.028$ & $0.083^{+0.030}_{-0.012}$ & 12 \\
NGC 6362 &  & $-1.092\pm 0.006$ & $<0.017$ & 28 \\
NGC 6366 &  & $-0.555\pm 0.028$ & $0.071^{+0.039}_{-0.016}$ & 29 \\
NGC 6388 &  & $-0.428\pm 0.008$ & $0.054^{+0.010}_{-0.004}$ & 2,30,31 \\
NGC 6397 &  & $-1.994\pm 0.004$ & $0.028^{+0.004}_{-0.002}$ & 1,2 \\
NGC 6402 & M 14 & $-1.130\pm 0.010$ & $0.053^{+0.011}_{-0.004}$ & 32 \\
NGC 6441 &  & $-0.334\pm 0.018$ & $0.079^{+0.016}_{-0.013}$ & 5 \\
NGC 6535 &  & $-1.963\pm 0.010$ & $0.035^{+0.012}_{-0.006}$ & 33 \\
NGC 6553 &  & $-0.151\pm 0.019$ & $<0.047$ & 34 \\
NGC 6569 &  & $-0.867\pm 0.014$ & $0.055^{+0.019}_{-0.005}$ & 35 \\
NGC 6626 &  & $-1.287\pm 0.016$ & $<0.075$ & 36 \\
NGC 6656 & M 22 & $-1.803\pm 0.015$ & $<0.132$ & 37,38 \\
NGC 6681 & M 70 & $-1.633\pm 0.014$ & $0.028^{+0.028}_{-0.001}$ & 39 \\
NGC 6715 & M 54 & $-1.559\pm 0.022$ & $0.183^{+0.022}_{-0.010}$ & 40 \\
NGC 6752 &  & $-1.583\pm 0.003$ & $0.034^{+0.003}_{-0.002}$ & 1,41,42 \\
NGC 6809 & M 55 & $-1.934\pm 0.003$ & $0.045^{+0.003}_{-0.002}$ & 1,2,3,43 \\
NGC 6838 & M 71 & $-0.736\pm 0.007$ & $0.039^{+0.007}_{-0.003}$ & 1,2,12 \\
NGC 6864 & M 75 & $-1.164\pm 0.018$ & $0.059^{+0.025}_{-0.008}$ & 44 \\
NGC 7078 & M 15 & $-2.287\pm 0.008$ & $0.053^{+0.006}_{-0.004}$ & 1,2,12 \\
NGC 7089 & M 2 & $-1.399\pm 0.011$ & $0.021^{+0.025}_{-0.008}$ & 12 \\
NGC 7099 & M 30 & $-2.356\pm 0.006$ & $0.037^{+0.006}_{-0.003}$ & 1,2 \\
Terzan 1 &  & $-1.263\pm 0.024$ & $0.037^{+0.047}_{-0.021}$ & 45 \\
Terzan 5 &  & $-0.092\pm 0.030$ & $0.295^{+0.030}_{-0.017}$ & 46 \\
Terzan 8 &  & $-2.255\pm 0.026$ & $0.098^{+0.037}_{-0.007}$ & 47 \\
\enddata
\tablerefs{(1) \citet{Carretta09-UVES}; (2) \citet{Carretta09-GIRAFFE}; (3) \citet{Wang17}; (4) \citet{Worley10}; (5) \citet{WillmanStrader12}; (6) \citet{Carretta15-N2808}; (7) \citet{Wang16}; (8) \citet{Schaeuble15}; (9) \citet{Carretta14-N4833}; (10) \citet{RoedererThompson15}; (11) \citet{Boberg16}; (12) \citet{Masseron19}; (13) \citet{Boberg15}; (14) \citet{Marino15-N5286}; (15) \citet{Carretta17-N5634}; (16) \citet{Bellazzini15}; (17) \citet{Mucciarelli13-N5694}; (18) \citet{Mucciarelli18-N5824}; (19) \citet{Roederer16}; (20) \citet{Johnson17-N5986}; (21) \citet{Carretta15-N6093}; (22) \citet{Bragaglia15}; (23) \citet{Carretta07-N6218}; (24) \citet{Johnson17-N6229}; (25) \citet{Lapenna15}; (26) \citet{Johnson17-N6273}; (27) \citet{Johnson15-N6273}; (28) \citet{Mucciarelli16-N6362}; (29) \citet{Johnson16}; (30) \citet{Carretta18-N6388}; (31) \citet{Carretta07-N6388}; (32) \citet{Johnson19-N6402}; (33) \citet{Bragaglia17}; (34) \citet{Tang17}; (35) \citet{Johnson18-N6569}; (36) \citet{Villanova17-M28}; (37) \citet{Marino11}; (38) \citet{Mucciarelli15-M22}; (39) \citet{OMalley17-N6681}; (40) \citet{Carretta10-M54}; (41) \citet{Carretta07-N6752}; (42) \citet{Yong13-N6752}; (43) \citet{Rain19}; (44) \citet{Kacharov13}; (45) \citet{Valenti15}; (46) \citet{Massari14-T5}; (47) \citet{Carretta14-T8}.}
\end{deluxetable}

\subsection{Notes on Individual Studies and Objects}\label{sec:individual-notes}

Objects whose analysis is worthy of note are listed below, as are clarifications regarding the treatment of some individual studies.

\subsubsection{Carretta et al.}
A large fraction of all data comes from the work of Carretta and collaborators using VLT FLAMES. For these studies \citep{Carretta07-N6752,Carretta07-N6218,Carretta07-N6388,Carretta09-GIRAFFE,Carretta09-UVES,Carretta10-M54,Carretta15-N6093,Carretta17-N5634,Carretta18-N6388,Carretta15-N2808,Bragaglia15,Bragaglia17}, we have adopted $\sigma_{\mathrm{lines}}/\sqrt{N_{\mathrm{lines}}}$ for the abundance uncertainties, and treated data taken with the UVES and GIRAFFE spectrographs as separate datasets.

\subsubsection{Johnson et al.}
There are a number of relevant studies by Johnson et al. \citep{Johnson15-N6273,Johnson16,Johnson17-N6229,Johnson17-N6273,Johnson17-N5986,Johnson18-N6569,Johnson19-N6402}. For some \citep{Johnson17-N5986}, the quoted error is $\sigma_{\mathrm{lines}}$, so these have been divided by $\sqrt{N_{\mathrm{lines}}}$;
for some \citep{Johnson16,Johnson17-N6229}, the quoted uncertainty already accounts for $N_{\mathrm{lines}}$ and can therefore be used unaltered;
for others \citep{Johnson18-N6569,Johnson19-N6402}, the individual values of $\sigma_{\mathrm{lines}}$ and $N_{\mathrm{lines}}$ are unpublished but were kindly provided by C. Johnson (private communication). In the case of NGC~6273 \citep{Johnson15-N6273,Johnson17-N6273}, the quoted errors include contributions from uncertainties in the atmospheric models, but the measured dispersion is large enough that using a line-only error makes no difference to the result.

\subsubsection{Masseron et al. (2019)}
\citet{Masseron19} used APOGEE near infrared spectra to observe 10 clusters. These are some of the only observations that are not in the optical, which may result in a systematic shift in the total metallicity but should not impact the metallicity spread. We have only used the RGB stars from this study. The $\sigma$ values in their Table~2 are actually $\sigma_{\mathrm{lines}}/\sqrt{N_{\mathrm{lines}}}$ (T. Masseron, private communication), and so they are adopted as the individual uncertainties.

\subsubsection{Mucciarelli et al.}
A number of VLT FLAMES studies by Mucciarelli and collaborators \citep{Mucciarelli13-N5694,Mucciarelli18-N5824,Bellazzini15} have quoted errors that are found to be overestimates with $\sigma_z \ll 1$. Dividing the quoted errors by $\sqrt{N_{\mathrm{lines}}}$ brings these studies into agreement with similar studies. UVES and GIRAFFE observations are treated as separate datasets, and only RGB stars have been used from these studies. NGC~6362 \citep{Mucciarelli16-N6362} and NGC~6656 \citep{Mucciarelli15-M22} are noted separately below.

\subsubsection{Schaeuble et al. (2015)}
For the \citet{Schaeuble15} study of NGC~4590, we only used RGB stars and assumed that $\sigma_{\mathrm{lines}}=0.13$ for all stars but calculated $N_{\mathrm{lines}}$ individually for each star based on their Table~3.

\subsubsection{Wang et al.}
Four clusters have been observed by \citet{Wang16,Wang17} with VLT FLAMES. We have treated their UVES and GIRAFFE datasets separately, and have used the NLTE \ion{Fe}{1} abundances. For NGC~2808, the random uncertainty is listed in \citet{Wang16}; for the other clusters only a total uncertainty that includes both random and systematic uncertainty was published, but the random uncertainties were kindly provided by Y. Wang (private communication).

\subsubsection{NGC 104 (47 Tuc)}
The GIRAFFE dataset from \citet{Carretta09-GIRAFFE} gives a lower value for $\sigma_0$ than the full sample ($\sigma_0 < 0.026)$ that is nominally inconsistent, while the UVES dataset from \citet{Carretta09-UVES} and GIRAFFE dataset from \citet{Wang17} give higher values ($\sigma_0 > 0.042$; there were not enough UVES stars in \citealt{Wang17} to use), but the magnitude of the discrepancy is not large.

\subsubsection{NGC 3201}
\citet{Simmerer13} found an extremely large abundance spread in this cluster using data from FLAMES/UVES on VLT and MIKE on Magellan, but \citet{Mucciarelli15-N3201} argue strongly that this is an artifact of their derived surface gravities; the \citet{Carretta09-GIRAFFE,Carretta09-UVES} data that we adopt show a much smaller dispersion.

\subsubsection{NGC 4833}
The \citet{Carretta14-N4833} and \citet{RoedererThompson15} data on NGC~4833 are nominally inconsistent; both Carretta GIRAFFE and UVES datasets favor $\sigma_0 < 0.018$ while the Roederer dataset on its own gives $\sigma_0=0.063^{+0.037}_{-0.011}$. The MCMC algorithm strongly favors the Carretta values due to the smaller individual errors and larger number of stars.

\subsubsection{NGC 5286}
The data for NGC~5286 come from \citet{Marino15-N5286}, with GIRAFFE and UVES datasets analyzed separately. We have adopted the $\sigma_{\mathrm{fit}}$ errors from their Tables 6 and 7. Adopting errors of $\sigma_{\mathrm{lines}}/\sqrt{N_{\mathrm{lines}}}$ gives identical results, while adopting their quoted $\sigma_{\mathrm{total}}$ results in a somewhat smaller dispersion of $\sigma_0=0.071^{+0.022}_{-0.005}$. 

\subsubsection{NGC 5904 (M 5)}
The individual datasets give significantly different results on their own, with \citet{Carretta09-GIRAFFE} preferring a small $\sigma_0 < 0.011$ and \citet{Masseron19} preferring a large $\sigma_0 > 0.067$. The MCMC algorithm splits the difference, and agrees with the result from \citet{Carretta09-UVES} on its own.

\subsubsection{NGC 6121 (M 4)}
The two GIRAFFE datasets give alternately low ($\sigma_0 < 0.019$; \citealp{Carretta09-GIRAFFE}) and high ($\sigma_0 > 0.074$; \citealp{Wang17}) dispersions. The MCMC algorithm splits the difference, and gives a value in agreement with both UVES datasets \citep{Carretta09-UVES,Wang17}.

\subsubsection{NGC 6362}
In \citet{Mucciarelli16-N6362}, the errors are described as a combination of line-based and atmosphere-based uncertainties, but the values don't match this description and give $\sigma_z \ll 1$, indicating that they are overestimated. We have taken their quoted typical values for both sources of uncertainty and added them in quadrature to use for all stars. For this GC, the result is quite sensitive to this choice; if we instead use only the line-based uncertainties, we find $\sigma_0 = 0.043^{+0.006}_{-0.003}$. Their [Na/Fe] distribution for the horizontal branch stars (which are not included in our analysis) is consistent with a unimodal distribution with the error values we have adopted.

\subsubsection{NGC 6397}
Although a large number of stars were observed by \citet{Lovisi12}, the errors are dramatically overestimated ($\sigma_z \ll 1$) and there is not sufficient information available to estimate them correctly. Taking the errors at face value, the Lovisi data on its own gives an upper limit $\sigma_0 < 0.036$, which is consistent with the value we derive and adopt from the Carretta data.

\subsubsection{NGC 6553}
Although \citet{Tang17} quote measurement errors of 0.03~dex using APOGEE, \citet{Holtzman15} finds that the internal consistency for iron abundances is 0.053~dex on average, so we have adopted that as the uncertainty for all stars.

\subsubsection{NGC 6626}
\citet{Villanova17-M28} tested the internal consistency of their precision, so we adopt their errors. However, they find a $4\sigma$ trend between their derived \feh\ values and the temperature of the star, which strongly suggests that there is a residual unphysical systematic within their data that is inflating the measured $\sigma_0$. The MCMC algorithm nominally gives $\sigma_0=0.054^{+0.021}_{-0.007}$, but due to this issue we have taken the top of this error bar as an upper limit on $\sigma_0$.

\subsubsection{NGC 6656 (M 22)}\label{sec:M22}
The data here come from \citet{Marino11}, as in \citet{WillmanStrader12}, but with the data from each telescope treated as a separate dataset and an error of 0.02 dex based on the equivalent width measurements, and from \citet{Mucciarelli15-M22} using their quoted errors that include equivalent width and atmospheric uncertainties. Mucciarelli et al. argue strongly that although the iron abundance based on \ion{Fe}{1} lines show a significant spread, the fact that the iron abundance from \ion{Fe}{2} lines do not means that the spread is an artifact, possibly due to non-LTE effects. We have therefore adopted the top of the error bar in our analysis as an upper limit on the true $\sigma_0$.

% \subsubsection{NGC 6715 (M 54)}
% The data from \citet{Mucciarelli17-M54} prefer a smaller dispersion with $\sigma_0 < 0.13$, but this could be related to the sharp \feh\ cutoff imposed between M54 and Sgr stars; the MCMC algorithm prefers the higher dispersion from \citet{Carretta10-M54} due to the smaller errors bars.

\subsubsection{NGC 6809 (M 55)}
For \citet{Rain19}, we have removed the likely variable star 1.

\subsubsection{NGC 7089 (M 2)}
\citet{Yong14-M2} found an extremely wide metallicity distribution in this cluster; however, \citet{Lardo16} reanalyzed these data and found that some of this spread is inconsistent with the analysis of \ion{Fe}{2} lines. Moreover, these studies have oversampled the metal-rich component, so the dispersion based on them is inflated relative to an unbiased sample. We have therefore not used either of these studies to derive $\sigma_0$, and just used \citet{Masseron19}.

\subsubsection{Terzan 5}
The uncertainties in \citet{Massari14-T5}, which include both measurement uncertainties and atmospheric model uncertainties, are taken verbatim, but the intrinsic dispersion in Terzan~5 is so large that the result is insensitive to any assumption about the size of the errors.

\subsection{Cross-Validation}\label{sec:xvalid}

\begin{figure*}
\plotone{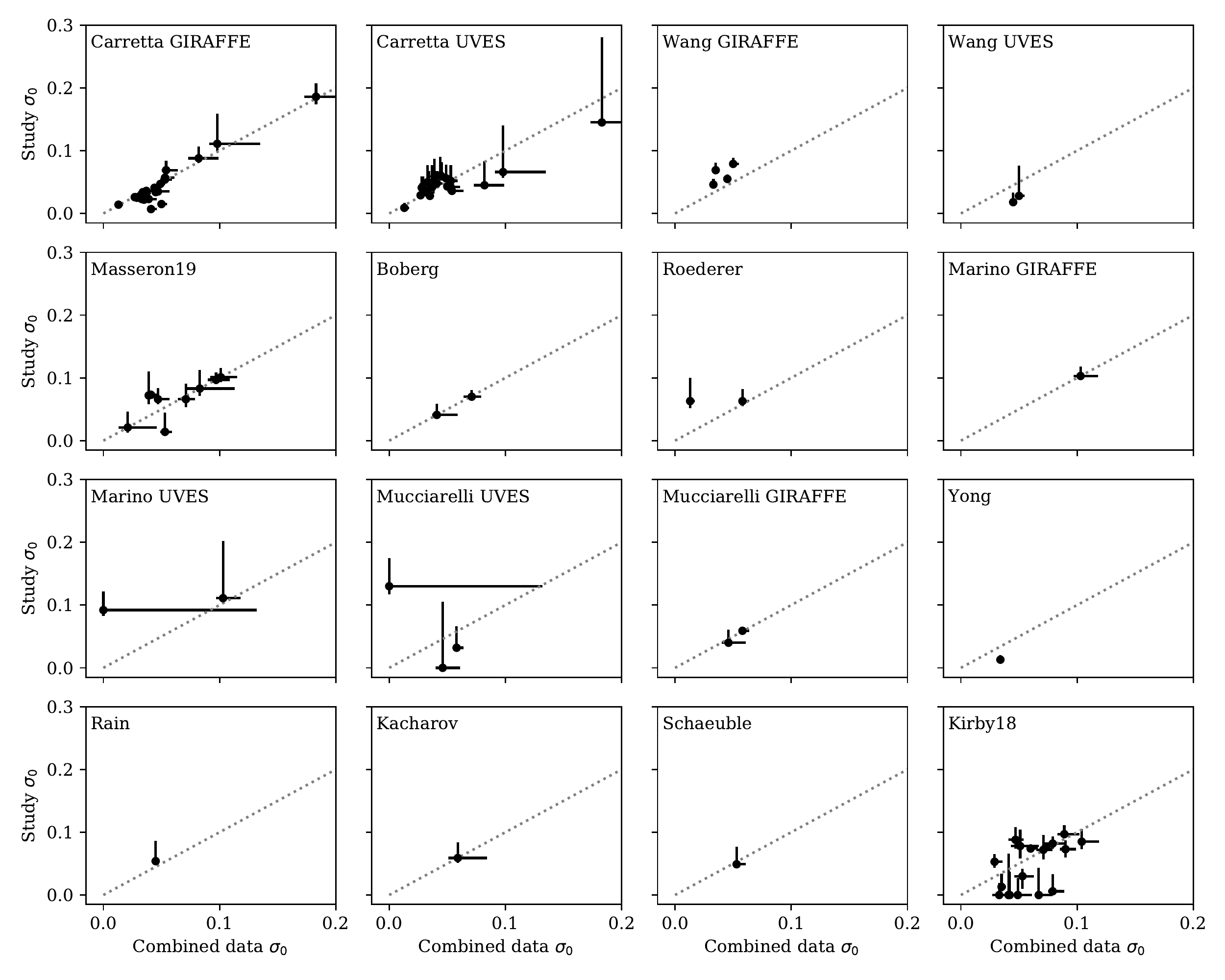}
\caption{Comparison between values of $\sigma_0$ for GCs derived from individual studies vs. from all available data for that cluster, for GCs that had at least two relevant datasets. Each author group/instrument combination is considered as a separate set of studies. The dashed line in each panel denotes the one-to-one line.
The Kirby18 panel shows what happens if we include data from \citet{Kirby18}, which uses spectral synthesis fitting rather than iron line equivalent widths (note that for this panel only, we have recomputed the ``Combined data $\sigma_0$'' to include the Kirby et al. data to enable a fair comparison).
All of the individual studies that we use have results that lie near the one-to-one line, indicating that our assessment of their errors and our MCMC method are robust; in contrast, the large scatter and lack of correlation in the Kirby18 plot demonstrates that spectral synthesis fitting measurements are insufficiently precise for determining abundance spreads.%
\label{fig:xvalid}}
\end{figure*}

\begin{figure*}
\plotone{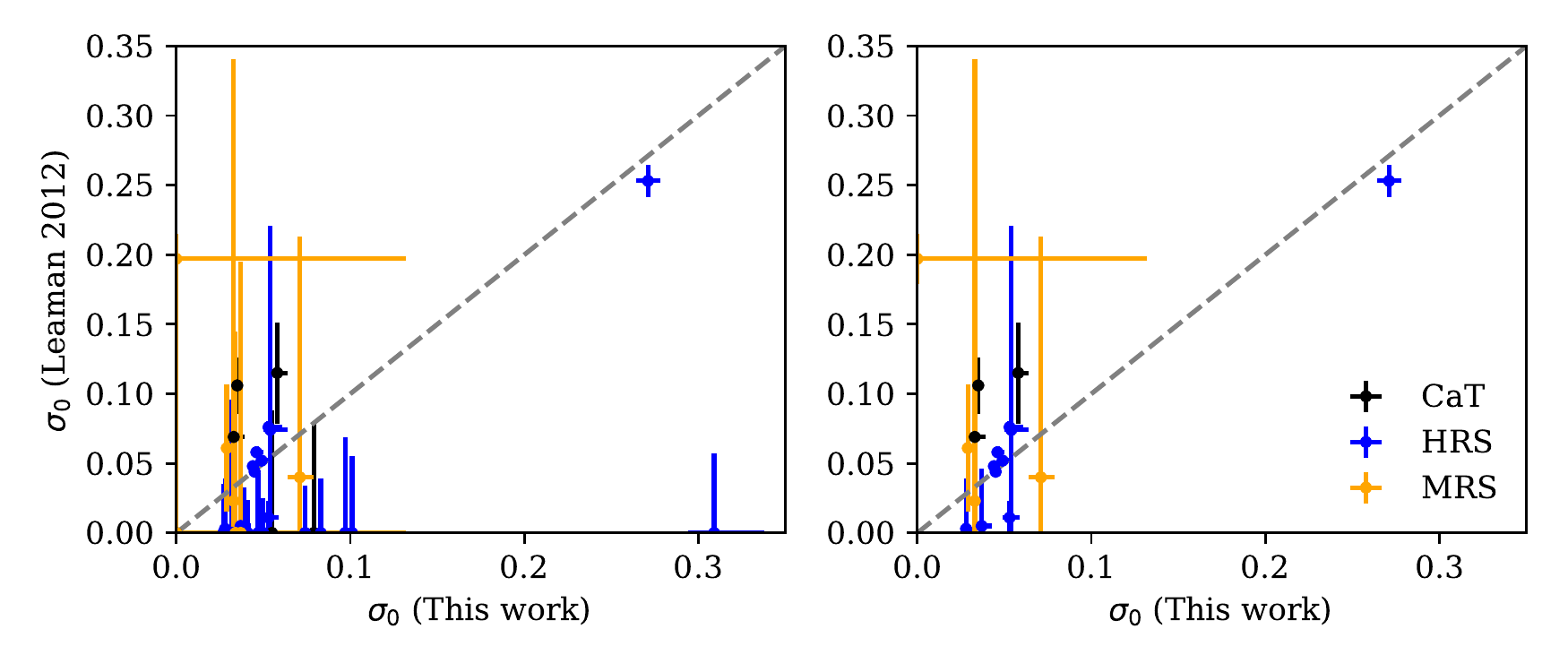}
\caption{Left: Comparison between values of $\sigma_0$ determined in this work vs. those derived by \citet{Leaman12} for clusters in common.
Clusters are colored by the observational method used in \citet{Leaman12}: Calcium Triplet (CaT), High Resolution Spectroscopy (HRS), or Medium Resolution Spectroscopy (MRS); in the present work, all data would count as High Resolution Spectroscopy. Right: As in the left panel but omitting the clusters listed as upper limits in \citet{Leaman12}}.\label{fig:leaman}
\end{figure*}

Because we have multiple datasets for many clusters, we can validate that our results are robust to the choice of datasets and our untangling of the random errors in many studies. We have grouped each author group/instrument combination together and compared the value of $\sigma_0$ we derive for each cluster if we \textit{only} use that dataset vs. our final result. Figure~\ref{fig:xvalid} shows the results of this cross-validation test. The vast majority of the data points for the studies that we use are consistent with the one-to-one line given their errors. In other words, 
\textbf{our final results using all available data are almost always consistent with each individual dataset when there are multiple datasets}. This test validates the robustness of the technique and decisions made regarding abundance uncertainties for individual studies.

When choosing relevant studies in the literature, we restricted ourselves to analyses that directly measured equivalent widths of iron lines to determine metallicities, which removed several large potential datasets, such as \citet{Dias16} and \citet{Kirby18}, that used spectral synthesis fitting. Our cross-validation test demonstrates why this is necessary; the Kirby18 panel in Figure~\ref{fig:xvalid} compares our values to those obtained using data from \citet{Kirby18}. The quoted uncertainties on individual stellar \feh\ measurements in \citet{Kirby18} are overestimates ($\sigma_z \ll 1$ for many clusters) due to their effective inclusion of both systematic and random errors; 0.07~dex has been removed in quadrature from the quoted errors, which brings $\sigma_z$ to  $\ga 1$ in most cases, though the comparison here is qualitatively unchanged if we use the quoted errors directly. Note also that \textbf{in this panel only} the Kirby et al. data is also used when calculating the ``Combined data $\sigma_0$'' to enable a fair comparison. This panel shows a large scatter with essentially no correlation, completely unlike the studies that measured individual iron lines. Therefore, we conclude that for the purpose of measuring intrinsic dispersions, only individual line equivalent width abundances can be calibrated sufficiently well.

We also compare our results to those of \citet{Leaman12}, who gathered observations of stellar abundances from a variety of sources and derived values of $\sigma_0$ (Figure~\ref{fig:leaman}); there are 34 clusters in common, with uncertainties kindly provided by R. Leaman (private communication). In the cases where we used the same data sources, our values are mostly (but not always) in agreement. Half of the clusters in common are nominally listed as upper limits in \citet{Leaman12} and are often inconsistent with measured values derived in this work, but these are rather cases where the errors were too large for Leaman to subtract the uncertainties on the individual abundance measurements rather than true upper limits on the value of $\sigma_0$. When these are omitted (as in the right panel of Figure~\ref{fig:leaman}), we see that the High Resolution Spectroscopy (HRS) measurements are mostly in agreement. The Medium Resolution Spectroscopy (MRS) measurements are also in agreement but this is largely because they have very large error, while the Calcium Triplet metallicities show no resemblance to ours. This demonstrates the critical importance of using resolved iron lines whose uncertainty can be accurately estimated in order to do this measurement, and of assessing those uncertainties in a homogeneous way.

\subsection{Overview of General Properties}\label{sec:properties}

\begin{figure}
\plotone{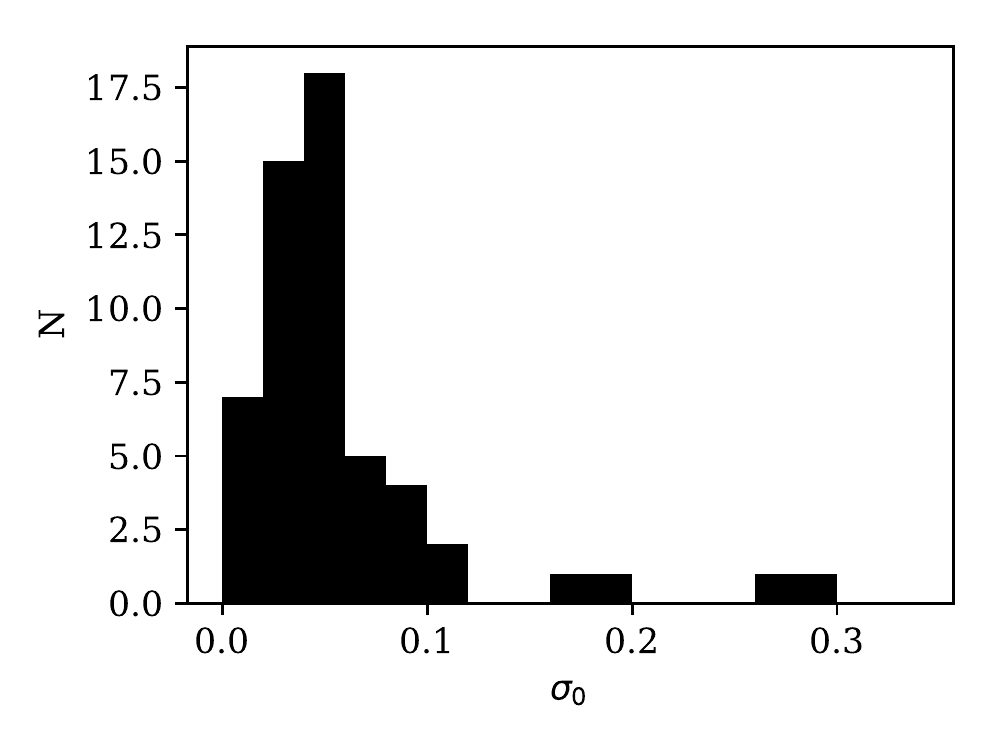}
\caption{Distribution of intrinsic iron dispersion $\sigma_0$ for the GCs in Table~\ref{table:sigma0}. The median value is $\sigma_0=0.045$~dex.\label{fig:sighist}}
\end{figure}

\begin{figure}
\plotone{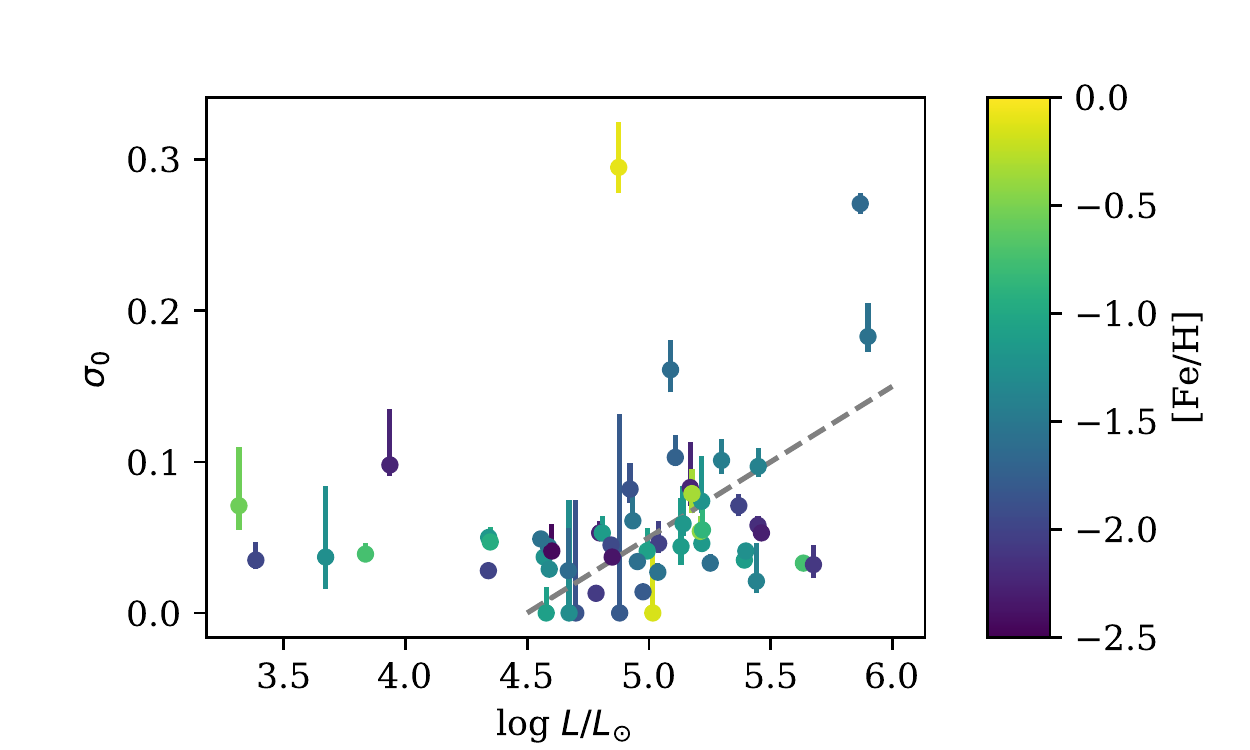}\\
\plotone{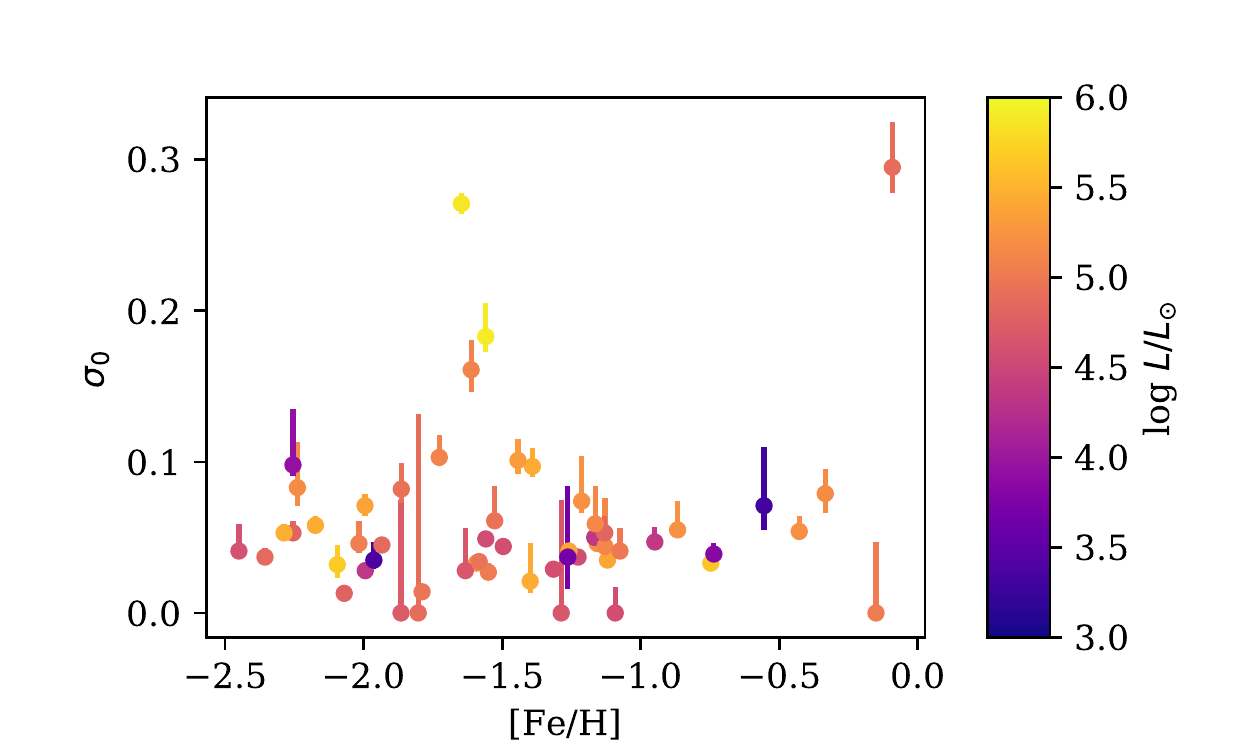}
\caption{Intrinsic iron dispersion $\sigma_0$ as a function of GC luminosity (top) and metallicity (bottom). Data points in the top panel are colored by metallicity, while data points in the bottom panel are colored by luminosity. High luminosity GCs with $L \ga 10^5~L_{\odot}$ have progressively larger values of $\sigma_0$; a dashed line of slope 0.1 is shown to guide the eye. No obvious trend is seen with metallicity.%
\label{fig:sigreln}}
\end{figure}

The basic properties of the GC sample are shown in Figures~\ref{fig:sighist} and \ref{fig:sigreln}.
Full analysis of the intrinsic metallicity dispersion will be given in Paper II (Bailin, in preparation) in the context of the \citet{Bailin18} clumpy self-enrichment model.

Most clusters have $\sigma_0 < 0.1$, with a median value of 0.045~dex and a mean value of 0.059~dex. This is certainly small (much smaller than the $\sim 1$~dex dispersions seen in abundances of intermediate-mass elements such as oxygen and sodium; e.g., \citealp{BastianLardo18} and references therein), but in the vast majority of cases are statistically distinguishable from zero (remember that the best fit value plotted here is also usually at the bottom of the error bar, so clusters that lie above 0 have robust detections of $\sigma_0>0$). Although GCs have often in the literature been separated into normal clusters with $\sigma_0=0$ and ``anomalous'' clusters with $\sigma_0>0$ \citep[e.g.][]{Marino15-N5286}, our results indicate that more nuance is required since most GCs demonstrably have small but non-zero dispersions. 
Previous authors have identified nine potentially anomalous GCs: NGC~1851, NGC~5139 ($\omega$~Cen), NGC~5286, NGC~5824, NGC~6656 (M~22), NGC~6715 (M~54), NGC~7089 (M~2), Terzan~5 \citep{Marino15-N5286}, and NGC~6273 (M~19) \citep{Johnson15-N6273}.
Out of those, we find that $\omega$~Cen, NGC~5286, possibly M~22 (see Section~\ref{sec:M22}), M~54, Terzan~5, and NGC~6273 do indeed have particularly large intrinsic iron dispersions with $\sigma_0 > 0.1$, while NGC~1851 and NGC~5824 have typical values of $\sigma_0\sim 0.05$ and M~2 has a quite small $\sigma_0=0.021$.
NGC~5272 and NGC~6205 also have relatively large values of $\sigma_0 \ga 0.1$ and are potentially interesting targets to look for other chemical anomalies, such as in \textit{s}-process elements.

We have compared $\sigma_0$ to the basic properties of each GC -- its luminosity and metallicity -- in Figure~\ref{fig:sigreln}. Luminosities are taken from \citet{Harris10}, an updated version of the \citet{Harris96} catalog; for GCs with Gaia distances in \citet{Helmi18}, the luminosity has been updated to account for the Gaia distance.
At $L \la 10^5~L_{\odot}$, there is no trend, but above this luminosity, the average dispersion increases with luminosity; a dashed line is shown to guide the eye, given by
\begin{equation}
\sigma_0 = 0.05 + 0.1 (\log L/L_{\odot} - 5).
\end{equation}
A trend between mass (and therefore luminosity) of the cluster and internal dispersion is expected from models of self-enrichment due to the increased depth of the gravitational potential when the cluster was forming \citep{Bailin18}.
There is no obvious trend between $\sigma_0$ and metallicity.

These results are consistent with and extend those of \citet{Carretta09-scale}, who analyzed spreads in their UVES and GIRAFFE data for 19 GCs. Although \citet{Carretta09-scale} did not remove the effects of abundance errors on all of the measured dispersions, their average dispersion of 0.048~dex is similar to our mean of 0.059~dex, but is smaller because their GC sample did not include \textit{any} of the clusters that have been identified as anomalous, and therefore did not include the clusters with particularly large values of $\sigma_0$. \citet{Carretta09-scale} also noted that iron spreads increase with luminosity, which we confirm.

\section{Conclusions}\label{sec:conclusions}

We have presented a technique to measure the intrinsic dispersion among datapoints with individual errors spanning multiple datasets that works even if the datasets have systematic offsets, as long as the measurement uncertainty of each individual datapoint can be estimated reasonably well. This technique could be applicable to a variety of problems that require measuring intrinsic dispersions from heterogeneous data.

We have used this technique to perform uniform homogeneous measurements of the intrinsic spread $\sigma_0$ in the iron abundance of RGB stars in 55 Milky Way globular clusters (GCs) for which sufficiently precise spectroscopic observations exist. Measuring individual iron lines is required in order to obtain precise enough metallicities, and accurate estimates of the \textbf{random} component of the uncertainties in metallicity measurements are also necessary. Although most GCs have small intrinsic dispersions with $\sigma_0 < 0.1$~dex, the dispersions are measurably non-zero, with a median of 0.045~dex. A few well-known objects, such as $\omega$~Cen, Terzan~5, M~54, and NGC~6273 have significantly higher dispersions, which is often interpreted as evidence for being stripped dwarf galaxy nuclei instead of true GCs. The values of $\sigma_0$ do not show an obvious correlation with GC metallicity, but a trend of larger $\sigma_0$ with higher luminosity appears above $L > 10^5~L_{\odot}$. Such a trend would be expected if the intrinsic dispersion comes at least partly from self-enrichment, where deeper gravitational potential wells during formation would lead to larger dispersions \citep{Bailin18}.

As a byproduct of this work, we have come up with the following best practices recommendations for quoting stellar abundance uncertainties from high spectral resolution data:
\begin{enumerate}
 \item List all components of uncertainty separately, with individual numbers for each star when the estimate can differ significantly between stars. In particular, we advise separating out (a) the line-to-line rms $\sigma_{\mathrm{lines}}$, (b) the number of lines used $N_{\mathrm{lines}}$, (c) systematic uncertainties (and, if multiple sources of systematic uncertainty are estimated, each source should be listed separately in addition to a combined number), (d) the sensitivity of abundances to each atmospheric parameter, (e) a total uncertainty due to atmospheric parameter uncertainty (see point 2), and (f) a total random uncertainty consisting of $\sigma_{\mathrm{lines}}/\sqrt{N_{\mathrm{lines}}}$ plus the uncertainty due to atmospheric parameters added in quadrature.
 \item When determining the uncertainty on abundances due to their dependence on atmospheric parameters, we recommend that the input random uncertainty on each atmospheric parameter be determined empirically from the data, and the source of this estimate be stated explicitly. A total uncertainty due to all atmospheric parameters should be calculated including covariance between parameters as in \citet{McWilliam95}.
\end{enumerate}
Providing this information will allow for the widest possible use of stellar abundance data by future researchers, since the different uncertainties affect different applications of the data differently.

\section*{Acknowledgements}

We express our sincere thanks to the authors of all of the studies whose data have been collected here for the hard work that went into those measurements. We particularly acknowledge the following authors for helpful discussions regarding the details of their data which were essential for making homogeneous comparable measurements: Christian Johnson, Evan Kirby, Ryan Leaman, Thomas Masseron, Ian Roederer, Jay Strader, Sandro Villanova, Yue Wang, Clare Worley.
We also thank Qin Wang for useful discussions, and the referee for helpful suggestions. This research has made use of NASA's Astrophysics Data System (ADS).

\software{Astropy \citep{astropy:2013}, emcee \citep{emcee}, Matplotlib \citep{matplotlib}, NumPy \citep{numpy}}

%%%%%%%%%%%%%%%%%%%%%%%%%%%%%%%%%%%%%%%%%%%%%%%%%%



\bibliographystyle{aasjournal}
\bibliography{gcsigmacat}

\begin{thebibliography}{}
\expandafter\ifx\csname natexlab\endcsname\relax\def\natexlab#1{#1}\fi
\providecommand{\url}[1]{\href{#1}{#1}}

\bibitem[{{Astropy Collaboration} {et~al.}(2013){Astropy Collaboration},
  {Robitaille}, {Tollerud}, {Greenfield}, {Droettboom}, {Bray}, {Aldcroft},
  {Davis}, {Ginsburg}, {Price-Whelan}, {Kerzendorf}, {Conley}, {Crighton},
  {Barbary}, {Muna}, {Ferguson}, {Grollier}, {Parikh}, {Nair}, {Unther},
  {Deil}, {Woillez}, {Conseil}, {Kramer}, {Turner}, {Singer}, {Fox}, {Weaver},
  {Zabalza}, {Edwards}, {Azalee Bostroem}, {Burke}, {Casey}, {Crawford},
  {Dencheva}, {Ely}, {Jenness}, {Labrie}, {Lim}, {Pierfederici}, {Pontzen},
  {Ptak}, {Refsdal}, {Servillat}, \& {Streicher}}]{astropy:2013}
{Astropy Collaboration}, {Robitaille}, T.~P., {Tollerud}, E.~J., {et~al.} 2013,
  \aap, 558, A33

\bibitem[{{Bailin}(2018)}]{Bailin18}
{Bailin}, J. 2018, \apj, 863, 99

\bibitem[{{Balbinot} \& {Gieles}(2018)}]{BalbinotGieles18}
{Balbinot}, E., \& {Gieles}, M. 2018, \mnras, 474, 2479

\bibitem[{{Bastian} \& {Lardo}(2018)}]{BastianLardo18}
{Bastian}, N., \& {Lardo}, C. 2018, \araa, 56, 83

\bibitem[{{Bekki} \& {Freeman}(2003)}]{Bekki03}
{Bekki}, K., \& {Freeman}, K.~C. 2003, \mnras, 346, L11

\bibitem[{{Bellazzini} {et~al.}(2015){Bellazzini}, {Mucciarelli}, {Sollima},
  {Catelan}, {Dalessandro}, {Correnti}, {D'Orazi}, {Cort{\'e}s}, \&
  {Amigo}}]{Bellazzini15}
{Bellazzini}, M., {Mucciarelli}, A., {Sollima}, A., {et~al.} 2015, \mnras, 446,
  3130

\bibitem[{{Boberg} {et~al.}(2015){Boberg}, {Friel}, \& {Vesperini}}]{Boberg15}
{Boberg}, O.~M., {Friel}, E.~D., \& {Vesperini}, E. 2015, \apj, 804, 109

\bibitem[{{Boberg} {et~al.}(2016){Boberg}, {Friel}, \& {Vesperini}}]{Boberg16}
---. 2016, \apj, 824, 5

\bibitem[{{Bragaglia} {et~al.}(2017){Bragaglia}, {Carretta}, {D'Orazi},
  {Sollima}, {Donati}, {Gratton}, \& {Lucatello}}]{Bragaglia17}
{Bragaglia}, A., {Carretta}, E., {D'Orazi}, V., {et~al.} 2017, \aap, 607, A44

\bibitem[{{Bragaglia} {et~al.}(2015){Bragaglia}, {Carretta}, {Sollima},
  {Donati}, {D'Orazi}, {Gratton}, {Lucatello}, \& {Sneden}}]{Bragaglia15}
{Bragaglia}, A., {Carretta}, E., {Sollima}, A., {et~al.} 2015, \aap, 583, A69

\bibitem[{{Carretta}(2015)}]{Carretta15-N2808}
{Carretta}, E. 2015, \apj, 810, 148

\bibitem[{{Carretta} \& {Bragaglia}(2018)}]{Carretta18-N6388}
{Carretta}, E., \& {Bragaglia}, A. 2018, \aap, 614, A109

\bibitem[{{Carretta} {et~al.}(2009{\natexlab{a}}){Carretta}, {Bragaglia},
  {Gratton}, {D'Orazi}, \& {Lucatello}}]{Carretta09-scale}
{Carretta}, E., {Bragaglia}, A., {Gratton}, R., {D'Orazi}, V., \& {Lucatello},
  S. 2009{\natexlab{a}}, \aap, 508, 695

\bibitem[{{Carretta} {et~al.}(2009{\natexlab{b}}){Carretta}, {Bragaglia},
  {Gratton}, \& {Lucatello}}]{Carretta09-UVES}
{Carretta}, E., {Bragaglia}, A., {Gratton}, R., \& {Lucatello}, S.
  2009{\natexlab{b}}, \aap, 505, 139

\bibitem[{{Carretta} {et~al.}(2014{\natexlab{a}}){Carretta}, {Bragaglia},
  {Gratton}, {D'Orazi}, {Lucatello}, \& {Sollima}}]{Carretta14-T8}
{Carretta}, E., {Bragaglia}, A., {Gratton}, R.~G., {et~al.} 2014{\natexlab{a}},
  \aap, 561, A87

\bibitem[{{Carretta} {et~al.}(2007{\natexlab{a}}){Carretta}, {Bragaglia},
  {Gratton}, {Lucatello}, \& {Momany}}]{Carretta07-N6752}
{Carretta}, E., {Bragaglia}, A., {Gratton}, R.~G., {Lucatello}, S., \&
  {Momany}, Y. 2007{\natexlab{a}}, \aap, 464, 927

\bibitem[{{Carretta} {et~al.}(2017){Carretta}, {Bragaglia}, {Lucatello},
  {D'Orazi}, {Gratton}, {Donati}, {Sollima}, \& {Sneden}}]{Carretta17-N5634}
{Carretta}, E., {Bragaglia}, A., {Lucatello}, S., {et~al.} 2017, \aap, 600,
  A118

\bibitem[{{Carretta} {et~al.}(2007{\natexlab{b}}){Carretta}, {Bragaglia},
  {Gratton}, {Catanzaro}, {Leone}, {Sabbi}, {Cassisi}, {Claudi}, {D'Antona},
  {Fran{\c c}ois}, {James}, \& {Piotto}}]{Carretta07-N6218}
{Carretta}, E., {Bragaglia}, A., {Gratton}, R.~G., {et~al.} 2007{\natexlab{b}},
  \aap, 464, 939

\bibitem[{{Carretta} {et~al.}(2007{\natexlab{c}}){Carretta}, {Bragaglia},
  {Gratton}, {Momany}, {Recio-Blanco}, {Cassisi}, {Fran{\c c}ois}, {James},
  {Lucatello}, \& {Moehler}}]{Carretta07-N6388}
---. 2007{\natexlab{c}}, \aap, 464, 967

\bibitem[{{Carretta} {et~al.}(2009{\natexlab{c}}){Carretta}, {Bragaglia},
  {Gratton}, {Lucatello}, {Catanzaro}, {Leone}, {Bellazzini}, {Claudi},
  {D'Orazi}, \& {Momany}}]{Carretta09-GIRAFFE}
---. 2009{\natexlab{c}}, \aap, 505, 117

\bibitem[{{Carretta} {et~al.}(2010){Carretta}, {Bragaglia}, {Gratton},
  {Lucatello}, {Bellazzini}, {Catanzaro}, {Leone}, {Momany}, {Piotto}, \&
  {D'Orazi}}]{Carretta10-M54}
---. 2010, \aap, 520, A95

\bibitem[{{Carretta} {et~al.}(2014{\natexlab{b}}){Carretta}, {Bragaglia},
  {Gratton}, {D'Orazi}, {Lucatello}, {Momany}, {Sollima}, {Bellazzini},
  {Catanzaro}, \& {Leone}}]{Carretta14-N4833}
---. 2014{\natexlab{b}}, \aap, 564, A60

\bibitem[{{Carretta} {et~al.}(2015){Carretta}, {Bragaglia}, {Gratton},
  {D'Orazi}, {Lucatello}, {Sollima}, {Momany}, {Catanzaro}, \&
  {Leone}}]{Carretta15-N6093}
---. 2015, \aap, 578, A116

\bibitem[{{Dias} {et~al.}(2016){Dias}, {Barbuy}, {Saviane}, {Held}, {Da Costa},
  {Ortolani}, {Gullieuszik}, \& {V{\'a}squez}}]{Dias16}
{Dias}, B., {Barbuy}, B., {Saviane}, I., {et~al.} 2016, \aap, 590, A9

\bibitem[{{Foreman-Mackey} {et~al.}(2013){Foreman-Mackey}, {Hogg}, {Lang}, \&
  {Goodman}}]{emcee}
{Foreman-Mackey}, D., {Hogg}, D.~W., {Lang}, D., \& {Goodman}, J. 2013, PASP,
  125, 306

\bibitem[{{Gaia Collaboration} {et~al.}(2018){Gaia Collaboration}, {Helmi},
  {van Leeuwen}, {McMillan}, {Massari}, {Antoja}, {Robin}, {Lindegren},
  {Bastian}, \& {Arenou}}]{Helmi18}
{Gaia Collaboration}, {Helmi}, A., {van Leeuwen}, F., {et~al.} 2018, \aap, 616,
  A12

\bibitem[{{Harris}(1996)}]{Harris96}
{Harris}, W.~E. 1996, \aj, 112, 1487

\bibitem[{{Harris}(2010)}]{Harris10}
---. 2010, arXiv e-prints, arXiv:1012.3224

\bibitem[{{Holtzman} {et~al.}(2015){Holtzman}, {Shetrone}, {Johnson}, {Allende
  Prieto}, {Anders}, {Andrews}, {Beers}, {Bizyaev}, {Blanton}, {Bovy},
  {Carrera}, {Chojnowski}, {Cunha}, {Eisenstein}, {Feuillet}, {Frinchaboy},
  {Galbraith-Frew}, {Garc{\'{\i}}a P{\'e}rez}, {Garc{\'{\i}}a-Hern{\'a}ndez},
  {Hasselquist}, {Hayden}, {Hearty}, {Ivans}, {Majewski}, {Martell},
  {Meszaros}, {Muna}, {Nidever}, {Nguyen}, {O'Connell}, {Pan}, {Pinsonneault},
  {Robin}, {Schiavon}, {Shane}, {Sobeck}, {Smith}, {Troup}, {Weinberg},
  {Wilson}, {Wood-Vasey}, {Zamora}, \& {Zasowski}}]{Holtzman15}
{Holtzman}, J.~A., {Shetrone}, M., {Johnson}, J.~A., {et~al.} 2015, \aj, 150,
  148

\bibitem[{{Hunter}(2007)}]{matplotlib}
{Hunter}, J.~D. 2007, Computing in Science and Engineering, 9, 90

\bibitem[{{Husser} {et~al.}(2016){Husser}, {Kamann}, {Dreizler}, {Wendt},
  {Wulff}, {Bacon}, {Wisotzki}, {Brinchmann}, {Weilbacher}, {Roth}, \&
  {Monreal-Ibero}}]{Husser16}
{Husser}, T.-O., {Kamann}, S., {Dreizler}, S., {et~al.} 2016, \aap, 588, A148

\bibitem[{{Johnson} {et~al.}(2019){Johnson}, {Caldwell}, {Michael Rich},
  {Mateo}, \& {Bailey}}]{Johnson19-N6402}
{Johnson}, C.~I., {Caldwell}, N., {Michael Rich}, R., {Mateo}, M., \& {Bailey},
  J.~I. 2019, \mnras, 485, 4311

\bibitem[{{Johnson} {et~al.}(2017{\natexlab{a}}){Johnson}, {Caldwell}, {Rich},
  {Mateo}, {Bailey}, {Clarkson}, {Olszewski}, \& {Walker}}]{Johnson17-N6273}
{Johnson}, C.~I., {Caldwell}, N., {Rich}, R.~M., {et~al.} 2017{\natexlab{a}},
  \apj, 836, 168

\bibitem[{{Johnson} {et~al.}(2017{\natexlab{b}}){Johnson}, {Caldwell}, {Rich},
  {Mateo}, {Bailey}, {Olszewski}, \& {Walker}}]{Johnson17-N5986}
---. 2017{\natexlab{b}}, \apj, 842, 24

\bibitem[{{Johnson} {et~al.}(2016){Johnson}, {Caldwell}, {Rich}, {Pilachowski},
  \& {Hsyu}}]{Johnson16}
{Johnson}, C.~I., {Caldwell}, N., {Rich}, R.~M., {Pilachowski}, C.~A., \&
  {Hsyu}, T. 2016, \aj, 152, 21

\bibitem[{{Johnson} {et~al.}(2017{\natexlab{c}}){Johnson}, {Caldwell}, {Rich},
  \& {Walker}}]{Johnson17-N6229}
{Johnson}, C.~I., {Caldwell}, N., {Rich}, R.~M., \& {Walker}, M.~G.
  2017{\natexlab{c}}, \aj, 154, 155

\bibitem[{{Johnson} {et~al.}(2018){Johnson}, {Rich}, {Caldwell}, {Mateo},
  {Bailey}, {Olszewski}, \& {Walker}}]{Johnson18-N6569}
{Johnson}, C.~I., {Rich}, R.~M., {Caldwell}, N., {et~al.} 2018, \aj, 155, 71

\bibitem[{{Johnson} {et~al.}(2015){Johnson}, {Rich}, {Pilachowski}, {Caldwell},
  {Mateo}, {Bailey}, \& {Crane}}]{Johnson15-N6273}
{Johnson}, C.~I., {Rich}, R.~M., {Pilachowski}, C.~A., {et~al.} 2015, \aj, 150,
  63

\bibitem[{{Kacharov} {et~al.}(2013){Kacharov}, {Koch}, \&
  {McWilliam}}]{Kacharov13}
{Kacharov}, N., {Koch}, A., \& {McWilliam}, A. 2013, \aap, 554, A81

\bibitem[{{Kirby} {et~al.}(2018){Kirby}, {Xie}, {Guo}, {Kovalev}, \&
  {Bergemann}}]{Kirby18}
{Kirby}, E.~N., {Xie}, J.~L., {Guo}, R., {Kovalev}, M., \& {Bergemann}, M.
  2018, \apjs, 237, 18

\bibitem[{{Lapenna} {et~al.}(2015){Lapenna}, {Mucciarelli}, {Ferraro},
  {Origlia}, {Lanzoni}, {Massari}, \& {Dalessandro}}]{Lapenna15}
{Lapenna}, E., {Mucciarelli}, A., {Ferraro}, F.~R., {et~al.} 2015, \apj, 813,
  97

\bibitem[{{Lardo} {et~al.}(2016){Lardo}, {Mucciarelli}, \& {Bastian}}]{Lardo16}
{Lardo}, C., {Mucciarelli}, A., \& {Bastian}, N. 2016, \mnras, 457, 51

\bibitem[{{Leaman}(2012)}]{Leaman12}
{Leaman}, R. 2012, \aj, 144, 183

\bibitem[{{Lovisi} {et~al.}(2012){Lovisi}, {Mucciarelli}, {Lanzoni}, {Ferraro},
  {Gratton}, {Dalessandro}, \& {Contreras Ramos}}]{Lovisi12}
{Lovisi}, L., {Mucciarelli}, A., {Lanzoni}, B., {et~al.} 2012, \apj, 754, 91

\bibitem[{{Marino} {et~al.}(2011){Marino}, {Sneden}, {Kraft}, {Wallerstein},
  {Norris}, {Da Costa}, {Milone}, {Ivans}, {Gonzalez}, {Fulbright}, {Hilker},
  {Piotto}, {Zoccali}, \& {Stetson}}]{Marino11}
{Marino}, A.~F., {Sneden}, C., {Kraft}, R.~P., {et~al.} 2011, \aap, 532, A8

\bibitem[{{Marino} {et~al.}(2015){Marino}, {Milone}, {Karakas}, {Casagrande},
  {Yong}, {Shingles}, {Da Costa}, {Norris}, {Stetson}, {Lind}, {Asplund},
  {Collet}, {Jerjen}, {Sbordone}, {Aparicio}, \& {Cassisi}}]{Marino15-N5286}
{Marino}, A.~F., {Milone}, A.~P., {Karakas}, A.~I., {et~al.} 2015, \mnras, 450,
  815

\bibitem[{{Massari} {et~al.}(2014){Massari}, {Mucciarelli}, {Ferraro},
  {Origlia}, {Rich}, {Lanzoni}, {Dalessandro}, {Valenti}, {Ibata}, {Lovisi},
  {Bellazzini}, \& {Reitzel}}]{Massari14-T5}
{Massari}, D., {Mucciarelli}, A., {Ferraro}, F.~R., {et~al.} 2014, \apj, 795,
  22

\bibitem[{{Masseron} {et~al.}(2019){Masseron}, {Garc{\'{\i}}a-Hern{\'a}ndez},
  {M{\'e}sz{\'a}ros}, {Zamora}, {Dell'Agli}, {Allende Prieto}, {Edvardsson},
  {Shetrone}, {Plez}, {Fern{\'a}ndez-Trincado}, {Cunha}, {J{\"o}nsson},
  {Geisler}, {Beers}, \& {Cohen}}]{Masseron19}
{Masseron}, T., {Garc{\'{\i}}a-Hern{\'a}ndez}, D.~A., {M{\'e}sz{\'a}ros}, S.,
  {et~al.} 2019, \aap, 622, A191

\bibitem[{{McWilliam} {et~al.}(1995){McWilliam}, {Preston}, {Sneden}, \&
  {Searle}}]{McWilliam95}
{McWilliam}, A., {Preston}, G.~W., {Sneden}, C., \& {Searle}, L. 1995, \aj,
  109, 2757

\bibitem[{{Mucciarelli} {et~al.}(2013){Mucciarelli}, {Bellazzini}, {Catelan},
  {Dalessandro}, {Amigo}, {Correnti}, {Cort{\'e}s}, \&
  {D'Orazi}}]{Mucciarelli13-N5694}
{Mucciarelli}, A., {Bellazzini}, M., {Catelan}, M., {et~al.} 2013, \mnras, 435,
  3667

\bibitem[{{Mucciarelli} {et~al.}(2018){Mucciarelli}, {Lapenna}, {Ferraro}, \&
  {Lanzoni}}]{Mucciarelli18-N5824}
{Mucciarelli}, A., {Lapenna}, E., {Ferraro}, F.~R., \& {Lanzoni}, B. 2018,
  \apj, 859, 75

\bibitem[{{Mucciarelli} {et~al.}(2019){Mucciarelli}, {Lapenna}, {Lardo},
  {Bonifacio}, {Ferraro}, \& {Lanzoni}}]{Mucciarelli19}
{Mucciarelli}, A., {Lapenna}, E., {Lardo}, C., {et~al.} 2019, \apj, 870, 124

\bibitem[{{Mucciarelli} {et~al.}(2015{\natexlab{a}}){Mucciarelli}, {Lapenna},
  {Massari}, {Ferraro}, \& {Lanzoni}}]{Mucciarelli15-N3201}
{Mucciarelli}, A., {Lapenna}, E., {Massari}, D., {Ferraro}, F.~R., \&
  {Lanzoni}, B. 2015{\natexlab{a}}, \apj, 801, 69

\bibitem[{{Mucciarelli} {et~al.}(2015{\natexlab{b}}){Mucciarelli}, {Lapenna},
  {Massari}, {Pancino}, {Stetson}, {Ferraro}, {Lanzoni}, \&
  {Lardo}}]{Mucciarelli15-M22}
{Mucciarelli}, A., {Lapenna}, E., {Massari}, D., {et~al.} 2015{\natexlab{b}},
  \apj, 809, 128

\bibitem[{{Mucciarelli} {et~al.}(2016){Mucciarelli}, {Dalessandro}, {Massari},
  {Bellazzini}, {Ferraro}, {Lanzoni}, {Lardo}, {Salaris}, \&
  {Cassisi}}]{Mucciarelli16-N6362}
{Mucciarelli}, A., {Dalessandro}, E., {Massari}, D., {et~al.} 2016, \apj, 824,
  73

\bibitem[{{Nordlander} {et~al.}(2012){Nordlander}, {Korn}, {Richard}, \&
  {Lind}}]{Nordlander12}
{Nordlander}, T., {Korn}, A.~J., {Richard}, O., \& {Lind}, K. 2012, \apj, 753,
  48

\bibitem[{{O'Malley} {et~al.}(2017){O'Malley}, {Knaizev}, {McWilliam}, \&
  {Chaboyer}}]{OMalley17-N6681}
{O'Malley}, E.~M., {Knaizev}, A., {McWilliam}, A., \& {Chaboyer}, B. 2017,
  \apj, 846, 23

\bibitem[{{Rain} {et~al.}(2019){Rain}, {Villanova}, {Mun{\~o}z}, \&
  {Valenzuela-Calderon}}]{Rain19}
{Rain}, M.~J., {Villanova}, S., {Mun{\~o}z}, C., \& {Valenzuela-Calderon}, C.
  2019, \mnras, 483, 1674

\bibitem[{{Renzini}(2013)}]{Renzini13}
{Renzini}, A. 2013, \memsai, 84, 162

\bibitem[{{Roederer} {et~al.}(2016){Roederer}, {Mateo}, {Bailey}, {Spencer},
  {Crane}, \& {Shectman}}]{Roederer16}
{Roederer}, I.~U., {Mateo}, M., {Bailey}, J.~I., {et~al.} 2016, \mnras, 455,
  2417

\bibitem[{{Roederer} \& {Thompson}(2015)}]{RoedererThompson15}
{Roederer}, I.~U., \& {Thompson}, I.~B. 2015, \mnras, 449, 3889

\bibitem[{{Schaeuble} {et~al.}(2015){Schaeuble}, {Preston}, {Sneden},
  {Thompson}, {Shectman}, \& {Burley}}]{Schaeuble15}
{Schaeuble}, M., {Preston}, G., {Sneden}, C., {et~al.} 2015, \aj, 149, 204

\bibitem[{{Simmerer} {et~al.}(2013){Simmerer}, {Ivans}, {Filler}, {Francois},
  {Charbonnel}, {Monier}, \& {James}}]{Simmerer13}
{Simmerer}, J., {Ivans}, I.~I., {Filler}, D., {et~al.} 2013, \apjl, 764, L7

\bibitem[{{Spite} {et~al.}(2016){Spite}, {Spite}, {Gallagher}, {Monaco},
  {Bonifacio}, {Caffau}, \& {Villanova}}]{Spite16}
{Spite}, M., {Spite}, F., {Gallagher}, A.~J., {et~al.} 2016, \aap, 594, A79

\bibitem[{{Tang} {et~al.}(2017){Tang}, {Cohen}, {Geisler}, {Schiavon},
  {Majewski}, {Villanova}, {Carrera}, {Zamora}, {Garcia-Hernandez}, {Shetrone},
  {Frinchaboy}, {Meza}, {Fern{\'a}ndez-Trincado}, {Mu{\~n}oz}, {Lin}, {Lane},
  {Nitschelm}, {Pan}, {Bizyaev}, {Oravetz}, \& {Simmons}}]{Tang17}
{Tang}, B., {Cohen}, R.~E., {Geisler}, D., {et~al.} 2017, \mnras, 465, 19

\bibitem[{{Valenti} {et~al.}(2015){Valenti}, {Origlia}, {Mucciarelli}, \&
  {Rich}}]{Valenti15}
{Valenti}, E., {Origlia}, L., {Mucciarelli}, A., \& {Rich}, R.~M. 2015, \aap,
  574, A80

\bibitem[{{van der Walt} {et~al.}(2011){van der Walt}, {Colbert}, \&
  {Varoquaux}}]{numpy}
{van der Walt}, S., {Colbert}, S.~C., \& {Varoquaux}, G. 2011, Computing in
  Science and Engineering, 13, 22

\bibitem[{{Villanova} {et~al.}(2017){Villanova}, {Moni Bidin}, {Mauro},
  {Munoz}, \& {Monaco}}]{Villanova17-M28}
{Villanova}, S., {Moni Bidin}, C., {Mauro}, F., {Munoz}, C., \& {Monaco}, L.
  2017, \mnras, 464, 2730

\bibitem[{{Wang} {et~al.}(2016){Wang}, {Primas}, {Charbonnel}, {Van der
  Swaelmen}, {Bono}, {Chantereau}, \& {Zhao}}]{Wang16}
{Wang}, Y., {Primas}, F., {Charbonnel}, C., {et~al.} 2016, \aap, 592, A66

\bibitem[{{Wang} {et~al.}(2017){Wang}, {Primas}, {Charbonnel}, {Van der
  Swaelmen}, {Bono}, {Chantereau}, \& {Zhao}}]{Wang17}
---. 2017, \aap, 607, A135

\bibitem[{{Willman} \& {Strader}(2012)}]{WillmanStrader12}
{Willman}, B., \& {Strader}, J. 2012, \aj, 144, 76

\bibitem[{{Worley} \& {Cottrell}(2010)}]{Worley10}
{Worley}, C.~C., \& {Cottrell}, P.~L. 2010, \mnras, 406, 2504

\bibitem[{{Yong} {et~al.}(2013){Yong}, {Mel{\'e}ndez}, {Grundahl}, {Roederer},
  {Norris}, {Milone}, {Marino}, {Coelho}, {McArthur}, {Lind}, {Collet}, \&
  {Asplund}}]{Yong13-N6752}
{Yong}, D., {Mel{\'e}ndez}, J., {Grundahl}, F., {et~al.} 2013, \mnras, 434,
  3542

\bibitem[{{Yong} {et~al.}(2014){Yong}, {Roederer}, {Grundahl}, {Da Costa},
  {Karakas}, {Norris}, {Aoki}, {Fishlock}, {Marino}, {Milone}, \&
  {Shingles}}]{Yong14-M2}
{Yong}, D., {Roederer}, I.~U., {Grundahl}, F., {et~al.} 2014, \mnras, 441, 3396

\end{thebibliography}

%%%%%%%%%%%%%%%%%%%%%%%%%%%%%%%%%%%%%%%%%%%%%%%%%%

%%%%%%%%%%%%%%%%% APPENDICES %%%%%%%%%%%%%%%%%%%%%

% \appendix

% \section{FIXME}

% FIXME.

%%%%%%%%%%%%%%%%%%%%%%%%%%%%%%%%%%%%%%%%%%%%%%%%%%

% Don't change these lines
\end{document}